\documentclass[aps,preprint,amsmath,amssymb,tightenlines,nofootinbib]{revtex4}
\usepackage{graphicx,color}% Include figure files
\usepackage{subfigure}
\def\be{\begin{eqnarray}}
\def\ed{\end{eqnarray}}

\newcommand{\ET}{\mbox{$\not \hspace{-0.10cm} E_T$ }}

\begin{document}

{\begin{flushright}{KIAS-P16053}
\end{flushright}}

\title{ Phenomenology of dark matter in chiral U(1)$_X$ dark sector}

\author{ P. Ko}
\email{pko@kias.re.kr}
\affiliation{School of Physics, KIAS, Seoul 02455, Korea}
\affiliation{Quantum Universe Center, KIAS, Seoul 02455, Korea}

\author{Takaaki Nomura}
\email{nomura@kias.re.kr}
\affiliation{School of Physics, KIAS, Seoul 02455, Korea}

\date{\today}

\begin{abstract}
We consider dark matter physics in a model for the dark sector with extra dark U(1)$_X$ gauge symmetry.  The dark sector is composed of exotic fermions that are charged under both dark U(1)$_X$ and the standard model SU(3)$_C \times$U(1)$_Y$ gauge groups, as well as standard model singlet complex scalars $\Phi$ and $X$ with nonzero U(1)$_X$ charge. In this model, there are two dark matter candidates$-$a scalar and a fermion$-$both of which are stabilized by accidental $Z_2$ symmetry. Their thermal relic density, and direct and indirect detection constraints are discussed in detail and we search for the parameter space of the model accommodating dark matter observations. We also discuss constraints from diphoton resonance searches associated with the scalar field which breaks the dark U(1)$_X$, in a way consistent with dark matter physics. In addition, implications for collider physics are discussed, focusing on the production cross section of the scalar boson.
\end{abstract}

\maketitle

\section{Introduction}

The standard model (SM) of particle physics has been very successful in describing 
experimental data at both low and high energies.   However, there are several remaining 
unsolved issues that require physics beyond the minimal SM.   Among these outstanding issue is 
explaining the nature of dark matter (DM), whose existence is confirmed through  
astronomical and cosmological observations.  

The existence of DM indicates a dark sector which is hidden from current experiments and observations.  
The nature of this dark sector is an open question. However,   as the SM is described 
by local gauge symmetries, it is plausible that the dark sector is also ruled by SM and/or hidden 
gauge symmetries.   In this sense, some particles in the dark sector can have charges of the SM 
gauge group which would induce interesting phenomena.  For example, the dark sector in 
supersymmetric extensions of the SM with $R$-parity conservation is composed of supersymmetric 
partners of SM particles, and most of them carry nonzero SM gauge charges. 
Moreover, we expect that these new particles in the dark sector may play a crucial role in 
explaining some anomalies observed in experiments. 

{ Using the LHC's 2015 experimental data, an excess of events in the diphoton channel around $m_{\gamma \gamma} \simeq 750$ GeV
was announced by both the ATLAS and CMS collaborations~\cite{Aaboud:2016tru, Khachatryan:2016hje,ATLAS:2016,CMS:2016owr} where the $\sim 5$ fb cross section for the process $pp \to R \to \gamma \gamma$
was indicated, with $R$ being a resonant state.
%To explain this diphoton excess, the production cross sections at $\sqrt{s}=13$ TeV are respectively 
%indicated for narrow width case as~\cite{Franceschini:2015kwy} 
%\begin{eqnarray}
%\sigma(pp\to R\to \gamma\gamma) = \left\{ \begin{array}{c} 
 %                     5.5 \pm 1.5 {\text{ fb}~~\text{ATLAS~\cite{ATLAS:2016, Aaboud:2016tru}}}\,, \\
 %                     4.8 \pm 2.1 {\text{ fb}~~\text{CMS~\cite{CMS:2016owr,Khachatryan:2016hje}\,, \ \ }} 
 %                                           \end{array}\right. \label{eq:data_di}
%\end{eqnarray}
%where $R$ stands for the diphoton resonance. 
%The best fit value of the width of $R$ by ATLAS is $\sim 45$ GeV, while a narrow width is preferred by CMS.
%$R$ should be either spin-0 or spin-2 particle, and we shall consider a scalar particle case ($R\equiv \phi$) with 750 GeV mass in this paper.  
To obtain the above cross section, $R$ is expected to couple with exotic particles which have electric charge and/or color in order to enhance the gluon fusion production of $R$ and its decay branching fraction 
into the diphoton mode.
Motivated by the 750 GeV diphoton excess, the present authors proposed a model for a dark sector with extra U(1)$_X$ dark gauge symmetry which is spontaneously broken, giving a massive dark photon $Z'$ decaying into SM fermions via kinetic mixing with SM gauge bosons~\cite{Ko:2016wce}.
Note that a number of authors have previously attempted to interpret this excess~\cite{Harigaya:2015ezk,Backovic:2015fnp,Angelescu:2015uiz,Nakai:2015ptz,Buttazzo:2015txu,DiChiara:2015vdm,Knapen:2015dap,Pilaftsis:2015ycr,Franceschini:2015kwy,Ellis:2015oso,Gupta:2015zzs,Kobakhidze:2015ldh,Falkowski:2015swt,Benbrik:2015fyz,Wang:2015kuj,Dev:2015isx,Allanach:2015ixl,Wang:2015omi,Chiang:2015tqz,Huang:2015svl,Ko:2016wce,Nomura:2016seu,Kanemura:2015bli,Cheung:2015cug,Nomura:2016fzs,Ko:2016lai,Ding:2015rxx}.
However, the new LHC data in 2016 disfavor the diphoton excess~\cite{CMS:2016crm,ATLAS:2016eeo}, where the upper limit of the cross section is given at $750$ GeV as $\sigma(pp \to R \to \gamma \gamma) \leq 1.21$ fb in narrow width approximation~\cite{ATLAS:2016eeo} taking into account 1 $\sigma$ fluctuation.
%To obtain the above cross section, $\phi$ is expected to couple with exotic particles which have electric 
%charge and/or color in order to enhance the gluon fusion production of $\phi$ and its decay branching fraction 
%into the diphoton mode.   Thus a dark sector with nonzero SM gauge charges is one of the interesting possibility to explain the diphoton excess.

Although motivated by the 750 GeV diphoton excess at first stage, we find our model is an interesting realization of a dark sector
with extra U(1)$_X$ dark gauge symmetry~\footnote{Some other models related to dark matter and extra U(1)$_X$ gauge symmetry are studied, e.~g.~in Refs.~\cite{Dudas:2013sia,Ducu:2015fda,Alves:2015mua,Alves:2015pea,Alves:2013tqa,Martinez:2015wrp,Martinez:2014rea,Martinez:2014ova,Baek:2014kna,Ko:2014nha,Ko:2014loa,Guo:2015lxa,Baek:2013qwa,Chiang:2013kqa}.}.  }
In this model, we introduce dark fermions which are vector-like under SU(3)$_C \times$U(1)$_Y$ gauge 
symmetry,  but chiral under U(1)$_X$, and U(1)$_X$ charged scalar fields $\Phi$ and $X$ to break the U(1) symmetry 
and to make charged/colored dark fermions decays into SM fermions and DM $X$.
Since the dark fermions are chiral, their masses are generated by the spontaneous U(1)$_X$ breaking 
due to the nonzero vacuum expectation value(VEV) of U(1)$_X$ charged scalar field $\Phi$ which is singlet 
under SM.
The signal of diphoton resonance is induced by scalar boson $\phi$ associated with $\Phi$ where its gluon fusion production and diphoton decay processes 
are induced through the dark fermion loop since the dark fermions couples to $\Phi$ and some of them carry color/electric charges.
Remarkably the Yukawa coupling between $\Phi$ and dark fermions are related to masses of dark fermions, which makes our model predictive.
Moreover an accidental $Z_2$ symmetry appears in our setup which provides stability of DM naturally.  
In our previous paper, the diphoton excess is mainly analyzed with limited parameter space but the phenomenology of DM is also interesting and worth for detailed analysis.
Thus, in this paper we carry out a detailed analysis of DM physics$-$including the relic density, and direct and indirect detection in the model$-$to explore the allowed parameter space.
Furthermore we also discuss compatibility with the current constraint from diphoton resonance search and implications for collider physics.

The paper is organized as follows.
In Sec. II, we review our model showing the particle contents and their mass spectra after spontaneous gauge 
symmetry breaking. 
We study the DM physics such as relic density, direct and indirect detection constraints searching for allowed parameter region, in Sec.III.
In Sec. IV, we discuss  the constraint from diphoton resonance search and implications for collider physics in the model.
We give the summary and discussion in Sec. V.

\section{The model}
\begin{center} 
\begin{table}[b]
%\begin{tiny}
\begin{tabular}{|c||c|c|c|c|c|c|c|c||c|c|}\hline\hline  
&\multicolumn{8}{c||}{Fermions} & \multicolumn{2}{c|}{Scalar} \\\hline
& ~$E_L$~ & ~$E_R^{}$~ & ~$N^{}_{L}$ ~ & ~$N_R$~  & ~$U_L$~ & ~$U_R$  & ~$D_L$  & ~$D_R$ & ~$\Phi$~ & ~$X$~ \\\hline 
SU(3) & $\bf{1}$ & $\bf{1}$  & $\bf{1}$ & $\bf{1}$ & $\bf{3}$ & $\bf{3}$  & $\bf{3}$   & $\bf{3}$ & $\bf{1}$   & $\bf{1}$ \\\hline 
SU(2) & $\bf{1}$ & $\bf{1}$  & $\bf{1}$ & $\bf{1}$ & $\bf{1}$ & $\bf{1}$  & $\bf{1}$   & $\bf{1}$ & $\bf{1}$   & $\bf{1}$ \\\hline 
U(1)$_Y$ & $-1$ & $-1$  & $0$ & $0$ & $\frac{2}{3}$ & $\frac{2}{3}$ & $-\frac{1}{3}$  & $-\frac{1}{3}$ & $0$ & $0$  \\\hline
U(1)$_X$ & $a$ & $-b$  & $-a$ & $b$ & $-a$ & $b$ & $a$  & $-b$ & $a+b$ & $a$  \\ \hline
\end{tabular}
\caption{Contents of new fermions and scalar fields and their charge assignments under the gauge 
symmetry SU(3)$\times$SU(2)$_L\times$U(1)$_Y\times$U(1)$_X$.  
 We consider three families of dark fermions.}
\label{tab:1}
% \end{tiny}
\end{table}
\end{center}
In this section we recapitulate our dark sector model proposed in Ref.~\cite{Ko:2016wce}.
We consider a dark sector with U(1)$_X$ dark gauge symmetry, new fermions carrying both SM 
SU(3)$_C \times$U(1)$_Y$ quantum numbers and U(1)$_X$ charges, and SM singlet complex scalar 
fields as summarized in Table.~\ref{tab:1}.
The new fermions are vector-like under the SM gauge symmetry but chiral under U(1)$_X$. 
The gauge anomalies from triangle loops are  canceled due to the appropriate U(1)$_X$ charge assignments.
The Yukawa interactions and the scalar potential which contain the new fields are given by
\begin{align}
L_{\rm Yukawa} =& y^E \bar E_L E_R \Phi + y^N \bar N_L N_R \Phi^\dagger 
+ y^U \bar U_L U_R \Phi^\dagger + y^D \bar D_L D_R \Phi \nonumber \\
&  + y^{Ee^i} \bar E_L e^i_R X + y^{Uu^i} \bar U_L u^i_R X^\dagger + y^{Dd^i} \bar D_L d^i_R X + h.c.,   \\
%\end{align}
%\begin{align}
V =& \mu^2 H^\dagger H + \lambda (H^\dagger H)^2 + \mu_\Phi^2 \Phi^\dagger \Phi 
+ \mu_X^2 X^\dagger  X \nonumber \\
&+ \lambda_\Phi (\Phi^\dagger \Phi)^2+ \lambda_X (X^\dagger X)^2   + \lambda_{H\Phi} 
(H^\dagger H)(\Phi^\dagger \Phi) \nonumber \\
& + \lambda_{HX} (H^\dagger H)(X^\dagger X)+ \lambda_{X\Phi} (X^\dagger X)(\Phi^\dagger \Phi).
\end{align}
where $H$ denotes the SM Higgs doublet field and the index $i$ denotes the SM fermion generation.
In this setup, there appears an accidental $Z_2$ symmetry: 
\[
X \to -X , \ \ \ F_L \to - F_L , \ \ \ F_R \to - F_R , 
\] 
which is not broken after U(1)$_X$ gauge symmetry breaking.
As a result the lightest $Z_2$ odd particle becomes stable and it can be DM candidate if it is a neutral one. 
Thus complex scalar $X$ and the lightest neutral dark fermion $N$ could be our DM candidates.
Note that this model is similar to the usual MSSM, except that the dark partners of the SM fermions are not 
scalars as in the MSSM, but fermions. And the complex scalar $X$ plays the role of the neutralino LSP 
(lightest supersymmetric particle). 

The gauge symmetry is broken after $H$ and $\Phi$ develop their nonzero VEVs: 
\begin{equation}
H= \left(\begin{array}{cc}
 G^+    \\
\frac{1}{\sqrt{2}} ( v+ h + iG^0)     
\end{array}
  \right)\,,  \quad
\Phi = \frac{1}{\sqrt{2}} (v_\phi + \phi + i G_\phi),
\end{equation}
where $G^\pm$, $G^0$ and $G_S$ are NG bosons which are absorbed by $W^\pm$, $Z$ and 
$Z'$ respectively.
The VEVs of the scalar fields are given approximately by 
\begin{equation}
v \simeq \sqrt{\frac{-\mu^2}{\lambda}}, \quad v_\phi \simeq \sqrt{\frac{-\mu_\Phi^2}{\lambda_\Phi}},
\end{equation}
where we assumed $\lambda_{H\Phi}$ is negligible so that the mixing between SM Higgs boson $h$ and $\phi$ is 
negligibly small to be consistent with the current Higgs data analysis~\cite{Chpoi:2013wga,Cheung:2015dta,Cheung:2015cug}. In this assumption, the mass of $h$ and $\phi$ are given by
\begin{equation}
m_h \simeq \sqrt{2 \lambda} v, \quad m_\phi \simeq \sqrt{2 \lambda_\Phi} v_\phi ,
\end{equation} 
where mass formula for $h$ is mostly the same as SM Higgs.
With the VEV of $\Phi$ the mass matrices of new fermions are given by 
\begin{equation}
M_F = \frac{y^F}{\sqrt{2}} v_\phi ,
\end{equation}
where $F=U,D,E,N$ and $M_F$ denotes mass of new fermion F and we have suppressed the family indices
for simplicity.

%%%%%%%%%%%%%%%%
We write the kinetic term for the gauge fields $\tilde B_\mu$ and $\tilde X_\mu$ which are from U(1)$_Y$ and U(1)$_X$ respectively, including kinetic mixing: 
\begin{align}
\mathcal{L}_{\text{kin}} =& -\frac{1}{4} W^a_{\mu \nu} W^{a \mu \nu} \nonumber \\
&- \frac{1}{4}(\tilde{B}_{\mu\nu},\tilde{X}_{\mu\nu})
\left(
\begin{array}{cc}
1 & s_\chi\\
s_\chi & 1
\end{array}
\right)
\left(
\begin{array}{c}
\tilde{B}^{\mu\nu}\\
\tilde{Z}^{'\mu\nu}
\end{array}\right), \label{kin} 
\end{align}
where $s_\chi \equiv \sin \chi$.
 Then we diagonalize the kinetic terms using the non-unitary transformation;
 \begin{align}
\left(
\begin{array}{c}
\tilde{B}_\mu\\
\tilde{X}_\mu
\end{array}\right)=
\left(
\begin{array}{cc}
1 & -t_\chi\\
0 & 1/t_\chi
\end{array}\right)
\left(
\begin{array}{c}
B_\mu\\
X_\mu
\end{array}\right) ~,
\end{align}
where $t_\chi = \tan \chi$.
After $\Phi$ and $H$ develop non-zero VEVs, we obtain the mass matrix for neutral gauge field approximately such that
\begin{equation}
\frac{1}{8} \left( \begin{array}{c} \tilde{Z} \\ X \end{array} \right)^T 
\left( \begin{array}{cc}  (g^2 +g'^2) v^2 & t_\chi g' \sqrt{g^2 + g'^2} v^2 \\ t_\chi g' \sqrt{g^2 + g'^2} v^2 &  4(a+b)^2 g_X^2 v_\phi^2 \end{array} \right)
 \left( \begin{array}{c} \tilde{Z} \\ X \end{array} \right).
\end{equation}
where $W^3_\mu = \cos \theta_W Z_\mu + \sin \theta_W A_\mu$ and $B_\mu = -\sin \theta_W Z_\mu + \cos \theta_W A_\mu$ are used.
In our analysis, we assume $\chi \ll 1$;   
actually the kinetic mixing parameter is experimentally limited roughly as $\chi \lesssim 10^{-2}-10^{-3}$ for $m_{Z'} \simeq O(100)$ GeV~\cite{Hook:2010tw, Andreas:2012mt,Jaeckel:2012yz}.
%\footnote{The upper bound on the kinetic mixing is roughly $\lesssim 0.01$ in the dark  photon mass range $m_{Z^\prime} \lesssim 350$ GeV considered in this letter~\cite{Jaeckel:2012yz}.}, 
With this assumption, the neutral gauge boson masses are approximated by 
\begin{equation}
m_Z^2 \simeq \frac{1}{4}(g^2+g'^2) v^2, \quad m_{Z'}^2 \simeq (a+b)^2 g_X^2 v_\phi^2.
\end{equation}
The mass eigenstates are also obtained as
\begin{align}
\left( \begin{array}{c} Z_\mu\\ Z'_\mu
\end{array}\right)=
\left(
\begin{array}{cc}
\cos \theta & -\sin \theta \\
\sin \theta & \cos \theta \end{array} \right)
\left( \begin{array}{c}
\tilde{Z}_\mu \\ X_\mu \end{array}\right) ~,
\end{align}
and the $Z-Z'$ mixing angle is given by
\begin{equation}
\tan 2 \theta \simeq \frac{g' \sqrt{g^2 + g'^2} v^2}{2 (m_Z^2 - m_{Z'}^2 ) } t_\chi,
\end{equation}
which is suppressed by $t_\chi$. 
Notice that $Z^\prime$ decays into the SM particles via the  kinetic mixing so that $\Gamma (Z^{'} )/m_{Z^\prime} \sim O(\chi^2) \lesssim 10^{-4}$. Therefore $Z^\prime$ would be a very narrow resonance.

%%%%%%%%%%%%%%%%

After U(1)$_X$ symmetry breaking, the interactions of physical scalar $\phi$ and $h$ are obtained from the Yukawa coupling and scalar potential such that
\begin{align}
{\cal L}_{\rm Yukawa} =& \frac{y^E}{\sqrt{2}} \bar E_L E_R \phi + \frac{y^N}{\sqrt{2}} \bar N_L N_R \phi 
+ \frac{y^U}{\sqrt{2}} \bar U_L U_R \phi + \frac{y^D}{\sqrt{2}} \bar D_L D_R \phi  \\
V \supset 
&  \frac{1}{4} \lambda_{H\Phi} (h h)(\phi \phi) + \frac{1}{2} \lambda_{H\Phi} v h (\phi \phi) +  \frac{1}{2} \lambda_{H\Phi} v_\phi \phi (h h) \nonumber \\
& + \frac{1}{2} \lambda_{HX} (h h)(X^\dagger X) + \lambda_{HX} v h(X^\dagger X)+ \frac{1}{2} \lambda_{X\Phi} (X^\dagger X)(\phi \phi) + \lambda_{X\Phi} v_\phi \phi (X^\dagger X).
\end{align}
Also the gauge interaction of $\phi$ is given by
\begin{equation}
{\cal L} \supset g_X^2 (a+b)^2 v_\phi \phi Z'^\mu Z'_\mu + \frac{1}{2} g_X^2 (a+b)^2 \phi \phi Z'^\mu Z'_\mu,
\end{equation}
where we took $\cos \theta \simeq 1$ since $\theta \ll 1$ as indicated above. In the following analysis we just apply $\theta = 0$.
The gauge interactions of DM candidates are given by
\begin{align}
\label{eq:gauge-int} 
{\cal L} \supset &  -i a g_X (\partial_\mu X^\dagger X - \partial_\mu X X^\dagger) Z'^\mu + a^2 g_X^2 Z'_\mu Z'^\mu X^\dagger X  \nonumber \\
&+ g_X ( a \bar N_L \gamma^\mu N_L  - b \bar N_R \gamma^\mu N_R ) Z'_\mu.
\end{align}
The gluon-gluon-$\phi$ coupling is induced by the new fermion loop, which is obtained as~\cite{Gunion:1989we}
  \begin{equation}
{\cal L}_{\phi gg} = \frac{\alpha_s}{8\pi } \left( \sum_{F=U,D} \frac{(a+b) \sqrt{2} g_X }{m_{Z'}} A_{1/2}(\tau_F)  \right) 
\phi G^{a\mu \nu}G^a_{\mu \nu} \,,\label{eq:LggS}
 \end{equation}
 where $A_{1/2}(\tau) =  2 \tau [1+(1-\tau) f(\tau)]$ with $f(\tau) = [\sin^{-1} (1/\sqrt{\tau})]$  for $\tau \geq 1$  and  $\tau_F \equiv 4 m_F^2/m_\phi^2$.
 Applying the relevant interactions, we can derive decay widths of $\phi$ into various channels, 
 which are summarized in the Appendix.

%%%%%%%%%%%%%%%%%%%%%%%%%%%%%%%%%%%%%%%%%%%%%%%%%%%%%%%%%%%%%%%%%%%%%%%%%%%%%%%%%%%%%%%%%%%%
%%%%%%%%%%%%%%%%%%%%%%%%%%%%%%%%%%%%%%%%%%%%%%%%%%%%%%%%%%%%%%%%%%%%%%%%%%%%%%%%%%%%%%%%%%%%

\section{Dark matter phenomenology}

In this section, we discuss phenomenology of dark matter in our model.
The dark matter of our model is the lightest neutral particle which is odd under accidental $Z_2$ symmetry; 
the candidates are $X$ and $N$.    In this work we consider two schemes;
\begin{align}
&(1):m_{DM} < m_{Z'}, \nonumber \\
&(2):m_{DM} > m_{Z'}. \nonumber
\end{align}
Then we focus on the procsses DM $\overline{\rm DM} \to$ gluons and DM $\overline{\rm DM} \to Z' Z'$ as the dominant DM annihilation processes for the schemes (1) and (2), respectively.   Notice that the annihilation processes DM $\overline{\rm DM} \to f_{SM} \overline{f_{SM}}$  are also possible via $\overline{F} f X$ 
Yukawa interactions described by Eq. (2).
Analysis of these interactions are already well studied by Refs.~\cite{Ibarra:2014qma,Giacchino:2015hvk} and we assume the contribution from the Yukawa contraction is small in our following analysis.

\subsection{Relic density}
%%%%%%%%%%%%%%%%%%%%%%%%%%%%%%%%%%%%%%%%%%%%%%%%%%%%%%%%%%%%%%%%%%
\begin{figure}[tb] 
\begin{center}
\includegraphics[width=70mm]{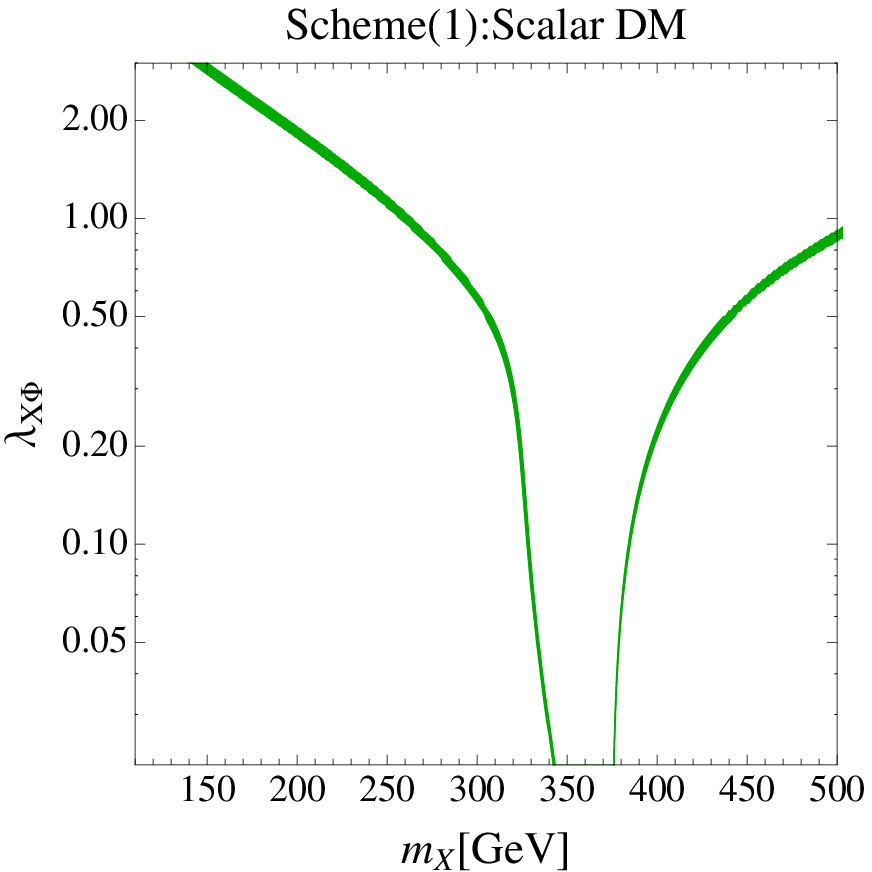}  
\includegraphics[width=70mm]{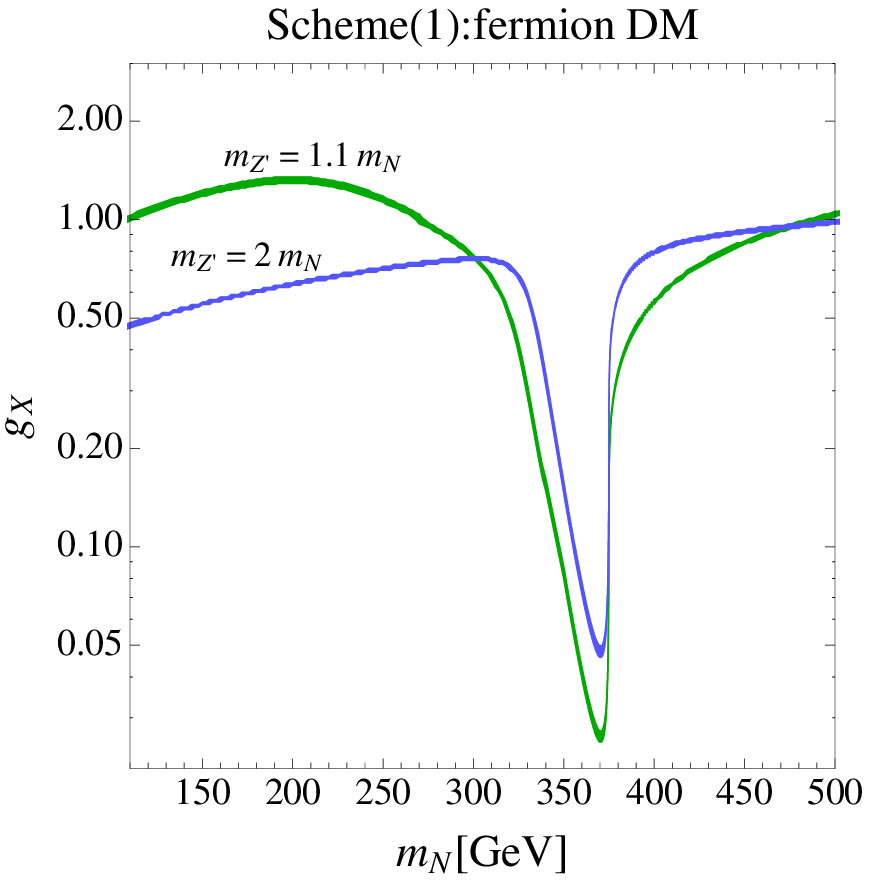}  
\caption{ The parameter region explaining the observed relic density of DM in the $m_X$-$\lambda_{X \Phi}$ 
and $M_N$-$g_X$ planes for scalar and fermionic DM cases, respectively,  in scheme (1). }
\label{fig:RDs1}
\end{center}
\end{figure}
%%%%%%%%%%%%%%%%%%%%%%%%%%%%%%%%%%%%%%%%%%%%%%%%%%%%%%%%%%
Here we estimate relic density of DM for both schemes (1) and (2) and search for the allowed parameter 
region of the model.   To reduce number of parameters in the analysis, we first fix some parameters in the 
model as follows:
\begin{align}
& M_E = 600 \, {\rm GeV}, \quad M_U = M_D = 800 \, {\rm GeV}, \quad m_{\phi} = 750 \, {\rm GeV}   \nonumber \\
& \lambda_{HX} = 0, \quad a \simeq 1, \quad b \simeq 1,  
\end{align}
where we assumed vanishing Higgs portal coupling\footnote{ The Higgs portal 
interaction would not affect our analysis much as long as  $\lambda_{XH} \lesssim 0.1$,
since DM annihilation cross section becomes less than $1/10$ of the cross section to provide 
observed relic density.  Therefore, we shall take $\lambda_{XH}=0$ hereafter for simplicity.} 
and $a \neq b$.    Then we assume $m_{\rm DM} < M_{E,U,D}$ to stabilize DM.  
In the following, we shall set the mass of $\phi$ to be 750 GeV since this mass point is well investigated due to the 750 GeV diphoton resonance, and we will discuss the constraints from recent data of diphoton resonance search in Sec. IV. 
We note that phenomenology would not much  change qualitatively  when we change the value 
of $m_\phi$, while some quantitative differences may appear for those processes where $\phi$ 
propagate in the s-channel; the position of the resonant region changes as 
$m_{DM} \simeq m_\phi/2$ and the required values of coupling constants that could explain 
thermal relic density of DM will get  larger (smaller) for heavier (lighter) $\phi$. 

For the scheme (1), the dominant DM annihilation processes are $X\overline{X} (N\overline{N}) \to \phi \to g g$ through the effective interaction Eq.~(\ref{eq:LggS}), 
which would be a good approximation as long as $m_{DM} < m_F$. 
The DM annihilation cross section can be obtained in non-relativistic approximation: 
\begin{align}
\label{eq:annihilationCX1}
( \sigma v_{\rm rel} )_{XX \to gg} \simeq & \left( \frac{m_{Z'}}{(a+b)g_X} \right)^2 \frac{\lambda_{X \Phi}^2}{(s - m_\phi^2)^2 + \Gamma_\phi^2 m^2_\phi} \frac{\Gamma(\phi \to g g)_{m_\phi = 2 m_{DM}}}{2 m_{DM}}, \\
\label{eq:annihilationCX2}
( \sigma v_{\rm rel} )_{NN \to gg} \simeq & 2 v_{\rm rel}^2 \frac{M_N^4}{m_{Z'}^2} \frac{(a+b)^2 g_X^2}{(s - m_\phi^2)^2 + \Gamma_\phi^2 m^2_\phi} \frac{\Gamma(\phi \to g g)_{m_\phi = 2 m_{DM}}}{2 m_{DM}},
\end{align}
where $s$ is center of mass total energy, $\Gamma_\phi$ is the total width of $\phi$, 
$\Gamma(\phi \to g g)_{m_\phi = 2 m_{DM}}$ is the width for the $\phi \to g g$ decay with 
$m_\phi = 2 m_{DM}$,  and we have used $v_\phi \simeq m_{Z'}/((a+b)g_X)$.
We note that $N\overline{N} \to gg$ does not have contribution from the S-wave, and the P-wave contribution 
would be dominant.  For $X\overline{X} \to gg$, the annihilation cross section is almost independent of $g_X$ 
and $m_{Z'}$  except for the resonant region around $m_X \sim m_\phi/2$ when we apply 
Eq.~(\ref{eq:width-gg}) to Eq.~(\ref{eq:annihilationCX1}).  
We thus scan $g_X$ in the region $\{0.1, 1.0\}$ and fix $m_{Z'} = 500$ GeV  for simplicity.   
On the other hand, the annihilation cross section depends on $g_X$ and $m_{Z'}$ for 
$N\overline{N} \to gg$ process.  Note that we fix $\lambda_{X\Phi} =0$ for fermion DM case since 
$\lambda_{X \Phi}$ is irrelevant parameter in  this case.
In this scheme, the total decay width of $\phi$ can be approximated as $\Gamma_\phi \simeq \Gamma 
(\phi \to gg) + \Gamma (\phi \to XX)$ since other modes are sufficiently small.
The relic density of DM is then obtained by solving the Boltzman equation. 
The approximated formula for the relic density is also given by~\cite{Gondolo:1990dk}, 
\begin{align}
\Omega h^2\approx \frac{1.07\times10^9  [{\rm GeV}]^{-1}} 
{g^{1/2}_* M_{\rm pl} \int_{x_f}^{\infty} \frac{dx}{x^2} \langle \sigma v_{\rm rel} \rangle_{\rm anni}},
%\int_{x_f}^\infty \left(\frac{a_{\rm eff}}{x^2}+6\frac{b_{\rm eff}}{x^3} \right)},
\label{eq:relic}
\end{align}
where $\langle \sigma v_{\rm rel} \rangle_{\rm anni}$ is thermal average of $\sigma v_{\rm rel}$ which is 
a function of  $x \equiv m_{DM}/T$ with temperature $T$, $x_f$ is $x$ at the freeze out temperature, 
$g_*$ is the total number of effective relativistic degrees of freedom at the time of freeze-out and 
$M_{\rm pl}=1.22\times 10^{19}[{\rm GeV}] $ is the Planck mass.
%To simplify the calculation, we adopt $x_f \simeq 25$ and $g_* \simeq 100$ which are typical values for DM mass of $O(100)$ GeV scale.
To estimate the relic density, we use { \tt micrOMEGAs 4.1.5}~\cite{Belanger:2014vza} where the Boltzmann equation is numerically solved by implementing relevant interactions for the annihilation processes.
Then, in our numerical analysis below, we set the approximated allowed region for the relic density to be~\cite{Ade:2013zuv}
\begin{align}
0.11\lesssim \Omega h^2\lesssim 0.13\label{eq:relicexp}.
\end{align}
In Fig.~\ref{fig:RDs1}, we show the parameter region which can account for DM thermal relic density for scalar 
and fermion DM cases in the left and right plots, respectively.
For the scalar DM case, we find that the required value of $\lambda_{X \Phi}$ becomes small at 
$m_X \sim m_\phi/2$ due to the resonant enhancement of the annihilation cross section.
For the fermion DM case, the dependence of $g_X$ on $m_{Z'}$ is not trivial and we show cases of 
$m_{Z'} = 1.1(2.0) m_N$ in the right plot of Fig.~\ref{fig:RDs1}. 

 Here we comment on the case with a non-negligible $h-\phi$ mixing $\alpha \sim O(0.1)$. In this case, DM annihilation processes $XX(NN) \to \phi \to W^+W^-/ZZ$ can be sizable via the scalar mixing effect if $m_{X(N)} \gtrsim O(100)$ GeV. Thus the parameter region 
satisfying thermal relic density of DM would change for scalar DM with $\lambda_{X\Phi} \neq 0$ 
and for fermion DM.  In particular, significant changes would appear for parameter region for 
the scheme (1) due to a small coupling of $\phi gg$ interaction (see Eq.~(\ref{eq:LggS})).

%%%%%%%%%%%%%%%%%%%%%%%%%%%%%%%%%%%%%%%%%%%%%%%%%%%%%%%%%%%%%%%%%%
\begin{figure}[tb] 
\begin{center}
\includegraphics[width=70mm]{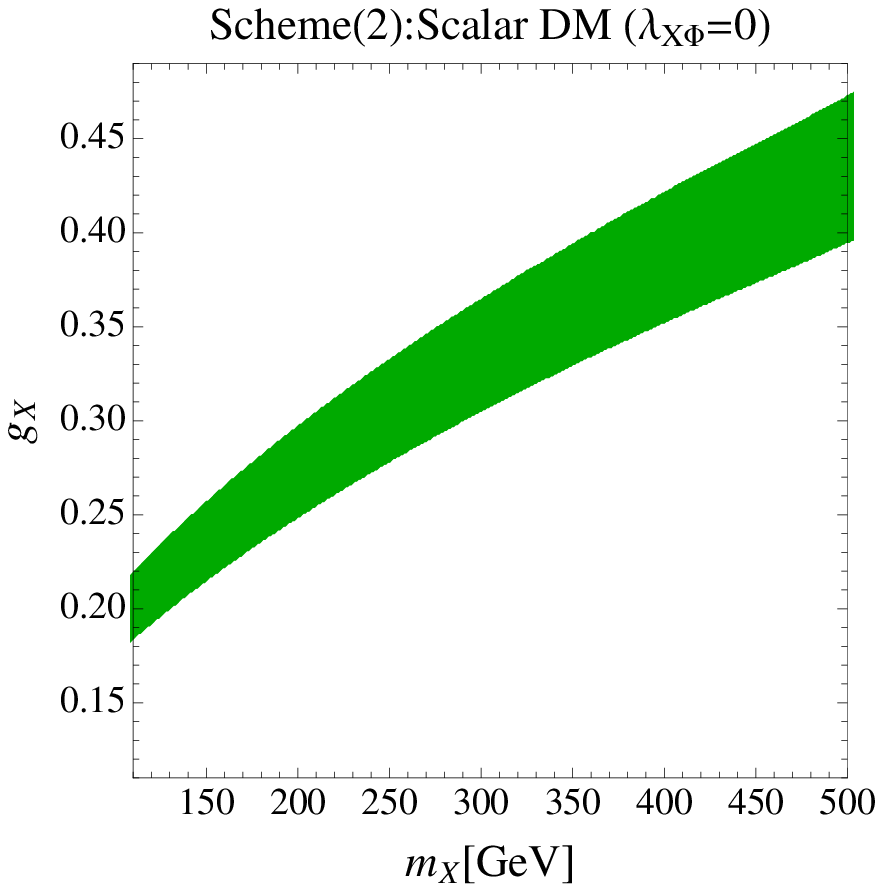} 
\includegraphics[width=70mm]{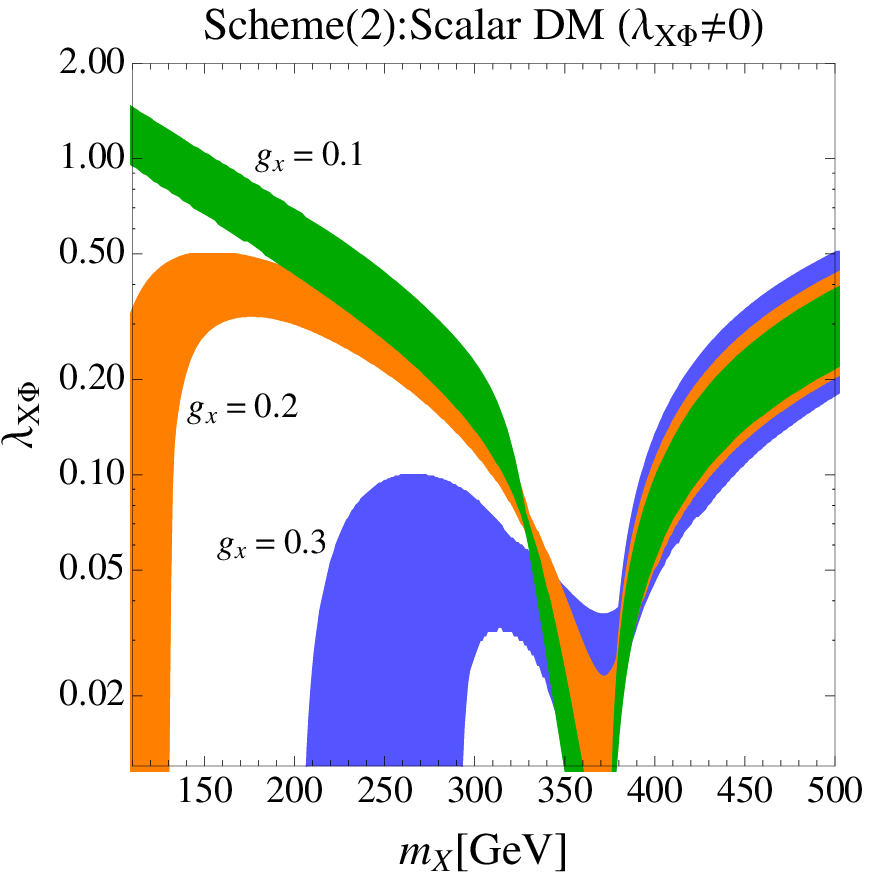} 
\caption{ The left (right) plot shows the parameter region explaining observed relic density of DM in the 
$g_X$-$m_X$ ($\lambda_{X \Phi}$-$m_X$) plane for scalar DM in scheme (2). }
\label{fig:RD1}
\end{center}
\end{figure}
%%%%%%%%%%%%%%%%%%%%%%%%%%%%%%%%%%%%%%%%%%%%%%%%%%%%%%%%%%
%%%%%%%%%%%%%%%%%%%%%%%%%%%%%%%%%%%%%%%%%%%%%%%%%%%%%%%%%%%%%%%%%%
\begin{figure}[tb] 
\begin{center}
\includegraphics[width=70mm]{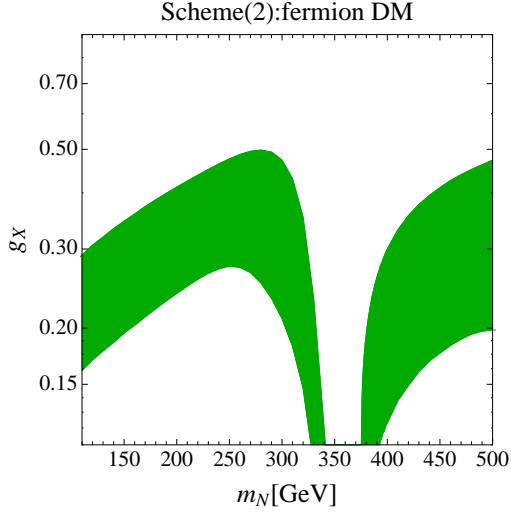}  
\caption{ The parameter region explaining observed relic density of DM in the $g_X$-$M_N$ plane 
for fermion DM in scheme (2). }
\label{fig:RD2}
\end{center}
\end{figure}
%%%%%%%%%%%%%%%%%%%%%%%%%%%%%%%%%%%%%%%%%%%%%%%%%%%%%%%%%%
For the scheme (2), we also numerically estimate thermal relic density of DM using { \tt micrOMEGAs 4.1.5 } to solve the Boltzmann equation by implementing relevant interactions which 
induce (co)annihilation processes of DM, $X \overline{X} (N \overline{N}) \to Z' Z'$.
The annihilation process $X^* X (\bar N N) \to Z' Z'$ are induced via gauge interaction Eq.~(\ref{eq:gauge-int}) 
and $\phi$ exchange in s-channel.   Thus coupling constants $g_X$, $\lambda_{X \Phi}$ and DM masses are 
relevant parameters in estimating the relic density of the DM.   We also run $Z'$ mass in the range of 
$0.3 \times m_X \leq m_{Z'} \leq 0.9 \times m_X$ to make the process kinematically allowed.
The left plot in Fig.~\ref{fig:RD1} shows the parameter region which explain relic density of the scalar DM in 
the $g_X$-$m_X$ plane where we take $\lambda_{X \Phi} =0$ and $M_N = 600$ GeV.
In this case, we find that $\sim0.15$ to $\sim 0.45$ gauge coupling can provide the observed relic density when DM mass is $\sim 100$ GeV to $\sim 500$ GeV. 
On the other hand, the right plot Fig.~\ref{fig:RD1} shows the parameter region in the $\lambda_{X\Phi}$-$m_X$ plane where we take $g_{X} =0.1$, $0.2$ and $0.3$ as reference values.
We notice that $\lambda_{X\Phi}$ should be very small for $m_X \sim m_\Phi/2$ since the annihilation cross 
section gets the Breit-Wigner enhancement.
For the case of fermion DM, we show the parameter region explaining the correct thermal relic density of 
DM in the $g_X$-$M_N$  plane in Fig.~\ref{fig:RD2}.    We find that s-channel process 
$N \overline{N} \to \phi \to Z' Z'$ provides dominant contribution to annihilation cross section for 
$M_N \sim m_{Z'}/2$.

\subsection{Direct detection}

\begin{figure}[tb] 
\begin{center}
\includegraphics[width=70mm]{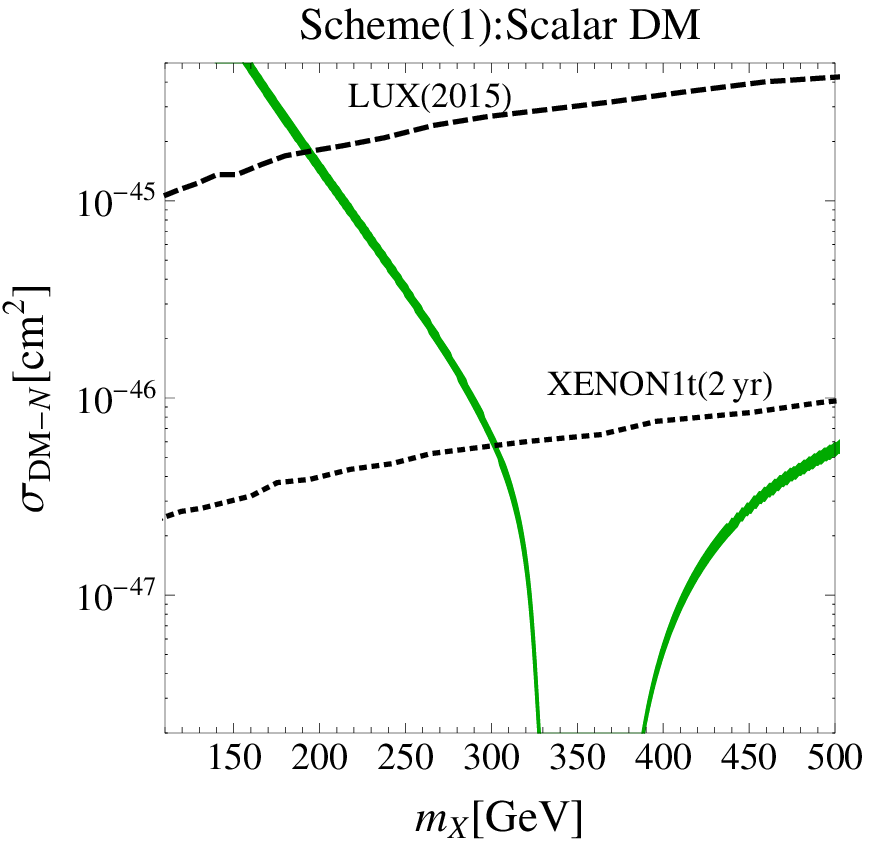} 
\includegraphics[width=70mm]{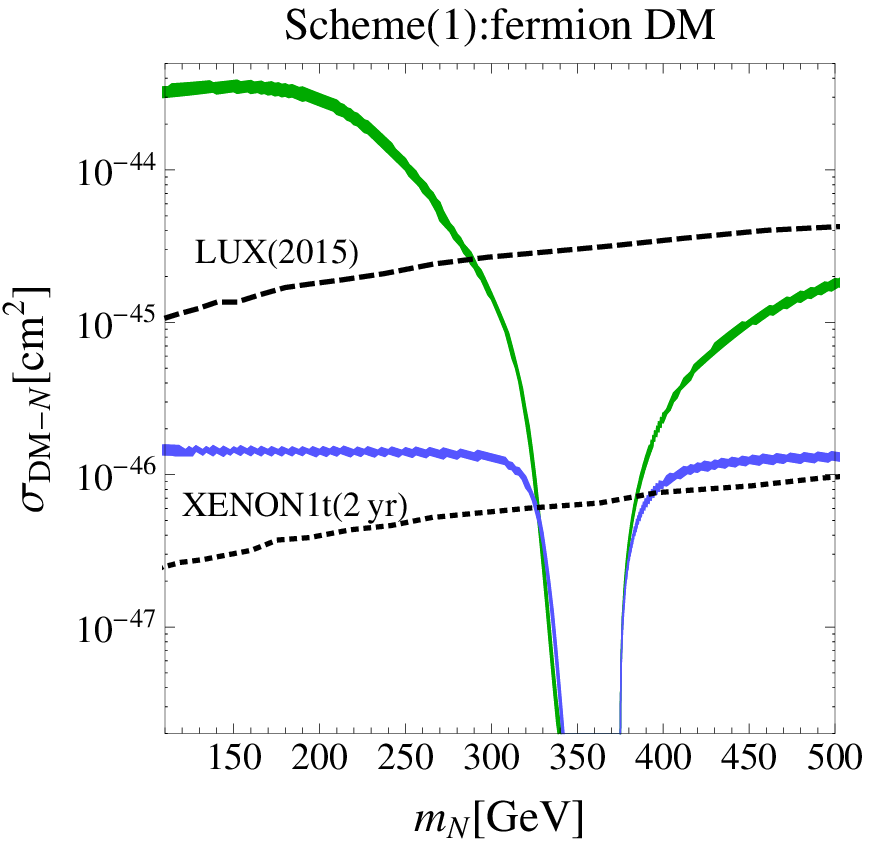} 
\caption{ The DM-nucleon scattering cross section as a function of DM mass for the scheme (1) where the parameters satisfying observed relic density in Fig.~\ref{fig:RDs1} are applied. The left and right plot correspond to scalar and fermionic DM cases respecteively. }
\label{fig:DDs1}
\end{center}
\end{figure}

\begin{figure}[tb] 
\begin{center}
\includegraphics[width=70mm]{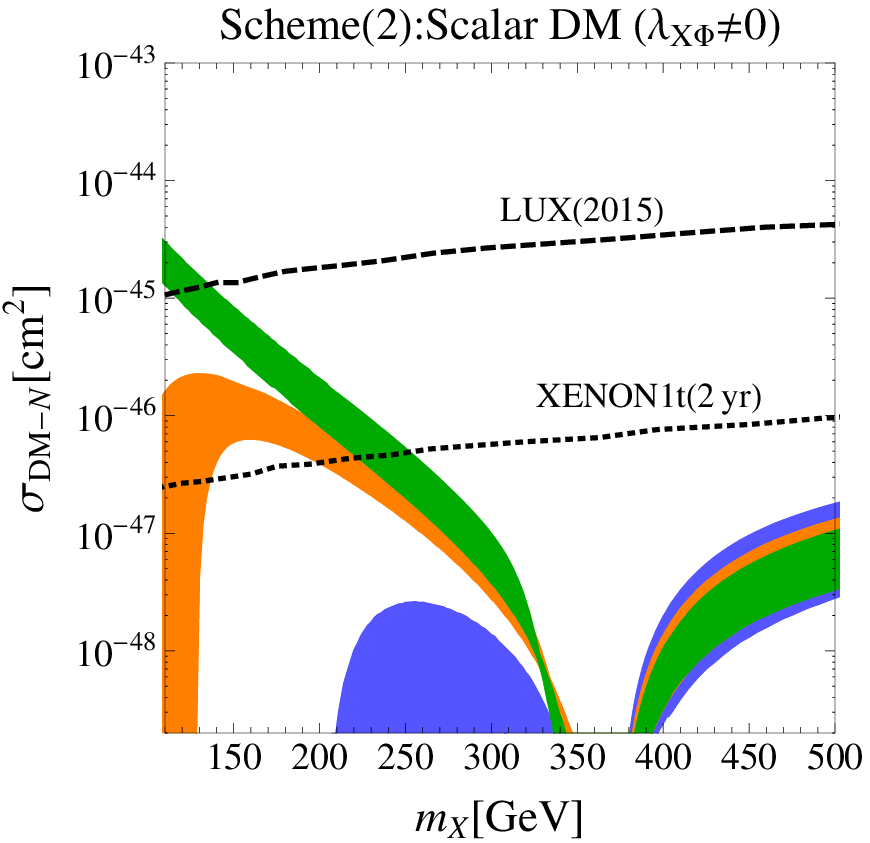} 
\includegraphics[width=70mm]{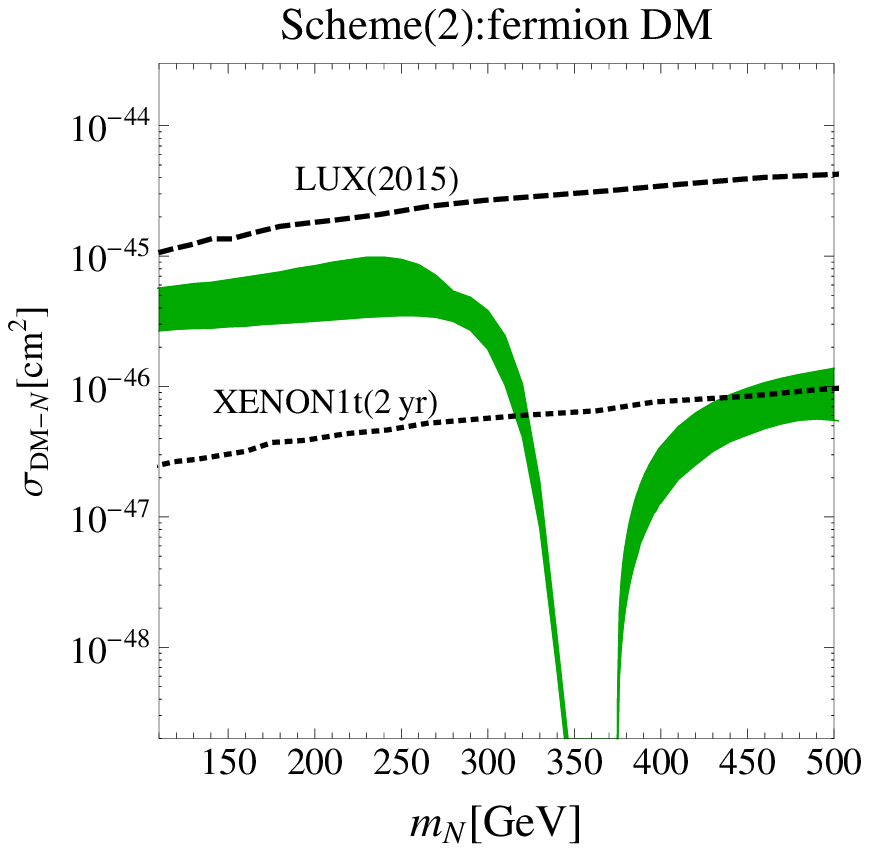} 
\caption{ The DM-nucleon scattering cross section as a function of DM mass for the scheme (2) where the parameters satisfying observed relic density in Fig.~\ref{fig:RD1} and \ref{fig:RD2} are applied. }
\label{fig:DD}
\end{center}
\end{figure}

In our mdoel,   DM-nucleon scattering occurs through the processes exchanging $h$, $\phi$ and 
$Z^\prime$ bosons.   The $Z^\prime$ exchange will provide small amplitude since it involves $Z-Z^\prime$ 
mixing which can be sufficiently small. 
Similarly the Higgs exchanging contribution can be made small enough if we take a small $\lambda_{HX}$. 
Therefore we shall focus on the $\phi$ exchange, which provide contribution to DM-nucleon scattering 
amplitude from  $\phi$-gluon-gluon coupling in Eq.~(\ref{eq:LggS}) and $\phi XX(\phi NN)$ coupling 
even if we ignore the $\phi-h$ mixing.     
The effective couplings for $XX(NN)$-$gg$ can be estimated as 
\begin{align}
{\cal L}_{XXGG} &= \frac{\alpha_S}{4 \pi} \left( \sum_{F=U,D} \frac{ \lambda_{X\Phi} }{m_{\phi}^2} A_{1/2}(\tau_F)  \right) X^\dagger X G^{a\mu \nu}G^a_{\mu \nu} \nonumber \\
& \equiv \frac{\alpha_S}{4 \pi} C^X_g X^\dagger X G^{a\mu \nu}G^a_{\mu \nu}, \\
{\cal L}_{NNGG} &= \frac{\alpha_S}{4 \pi} \left( \sum_{F=U,D} \frac{ 2 \sqrt{2} g_X^2 M_N}{m_{\phi}^2 m_{Z'}^2} A_{1/2}(\tau_F)  \right) \bar N N G^{a\mu \nu}G^a_{\mu \nu} \nonumber \\
& \equiv \frac{\alpha_S}{4 \pi} C^N_g \bar N N G^{a\mu \nu}G^a_{\mu \nu}.
\end{align}
The spin-independent $X(N)$-nucleon scattering cross section is obtained as~\cite{Giacchino:2015hvk} 
\begin{align}
\sigma_{\rm SI}^X &= \frac{m_{Nu}^2}{\pi (m_X + m_{Nu})^2} (f_{Nu}^X)^2 \\
\sigma_{\rm SI}^N &= \frac{2 m_N^2 m_{Nu}^2}{\pi (m_N + m_{Nu})^2} (f_{Nu}^N)^2 \\
\frac{f_{Nu}^{X(N)}}{m_{Nu}} &= - \frac{2}{9} C_g f^{(Nu)}_{T_G}
\end{align}
where $m_{Nu}$ is the nucleon mass and $f^{(Nu)}_{T_G}$ is the mass fraction of gluonic operators in the nucleon mass.  For the numerical values for these parameters, we adopt values in Ref.~\cite{Hisano:2015bma}. 
In Figs.~\ref{fig:DDs1} and \ref{fig:DD} we show the spin independent DM-nucleon scattering cross section 
where the allowed parameter regions from the relic density estimation are applied.
In scheme (1), we find that parameter region with $m_{X} \lesssim 200$ GeV is excluded for scalar DM and $m_N \lesssim 300$ GeV is excluded for fermion DM with $m_{Z'} = 1.1 m_N$,  by the current constraints of LUX experiment~\cite{Akerib:2013tjd, Akerib:2015rjg}.
Except for the resonant region, most of the parameter region can be tested in future direct detection experiment such as XENON 1t~\cite{Aprile:2015uzo}.
In scheme (2), we find that only parameter region with small $g_X$ and $m_X \lesssim 150$ GeV is constrained by the LUX data.
The other region will be explored by the future experiments.

%%%%%%%%%%%%%%%%%%%%%%%%%%%%%%%%%%%%%%%%%%%%%%%%%%%%%%%%%%%%%%%%%%
%%%%%%%%%%%%%%%%%%%%%%%%%%%%%%%%%%%%%%%%%%%%%%%%%%%%%%%%%%
%%%%%%%%%%%%%%%%%%%%%%%%%%%%%%%%%%%%%%%%%%%%%%%%%%%%%%%%%%%%%%%%%%
%%%%%%%%%%%%%%%%%%%%%%%%%%%%%%%%%%%%%%%%%%%%%%%%%%%%%%%%%%

\subsection{Indirect detection}

Here we discuss indirect detection of DM. 
In our model, DM pair annihilate dominantly into $gg$ and $Z' Z'$ in the schemes (1) and (2) respectively, 
and $Z'$ will further decay into SM fermions via kinetic mixing.

\begin{figure}[tb] 
\begin{center}
\includegraphics[width=80mm]{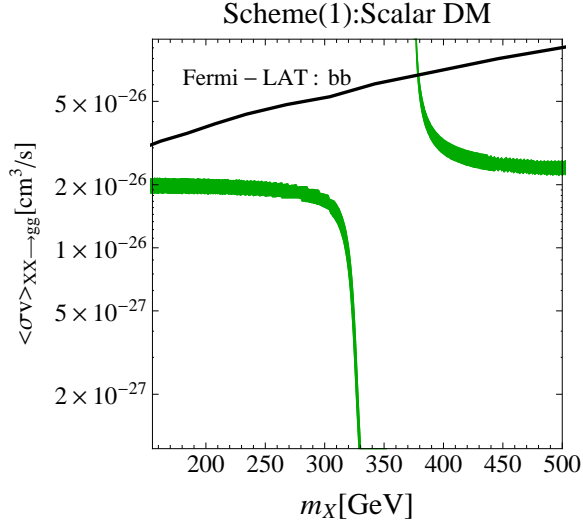} 
\caption{ The thermally averaged cross section for $XX \to gg$ in current Universe where the parameter region in Fig.~\ref{fig:RDs1} is applied. The black solid line indicate the current limit of the cross section for $DM DM \to b \bar b$ annihilation mode from Fermi-LAT}
\label{fig:IDs1}
\end{center}
\end{figure}

For the scheme (1) the annihilation cross section for $XX(NN) \to gg$ at the current Universe are estimated using { \tt micrOMEGAs 4.1.5 } where parameter sets providing observed relic density are applied as inputs.
In the case of scalar DM,  we obtain the thermally averaged cross section $\langle \sigma v \rangle_{XX \to gg}$ shown in Fig.~\ref{fig:IDs1} where the colored region correspond to the parameter space in the left plot of Fig.~\ref{fig:RDs1}.
Here we compared the cross section with current limit of the cross section for $DM DM \to b \bar b$ from Fermi-LAT~\cite{Ackermann:2015zua} where limit for $gg$ mode is indicated to be slightly weaker than $b \bar b$ mode~\cite{Huang:2015svl}.
Thus the parameter region $m_X \simeq m_\phi/2$ is excluded due to resonant enhancement while other regions are allowed.
In the case of fermion DM, we find that the current thermally averaged annihilation cross section is much smaller than the constraint by Fermi-LAT since $N\bar N \to \phi \to gg$ process is $t-$channel one.
Therefore here we omit the plot for the cross section for fermion DM. 

\begin{figure}[tb] 
\begin{center}
\includegraphics[width=80mm]{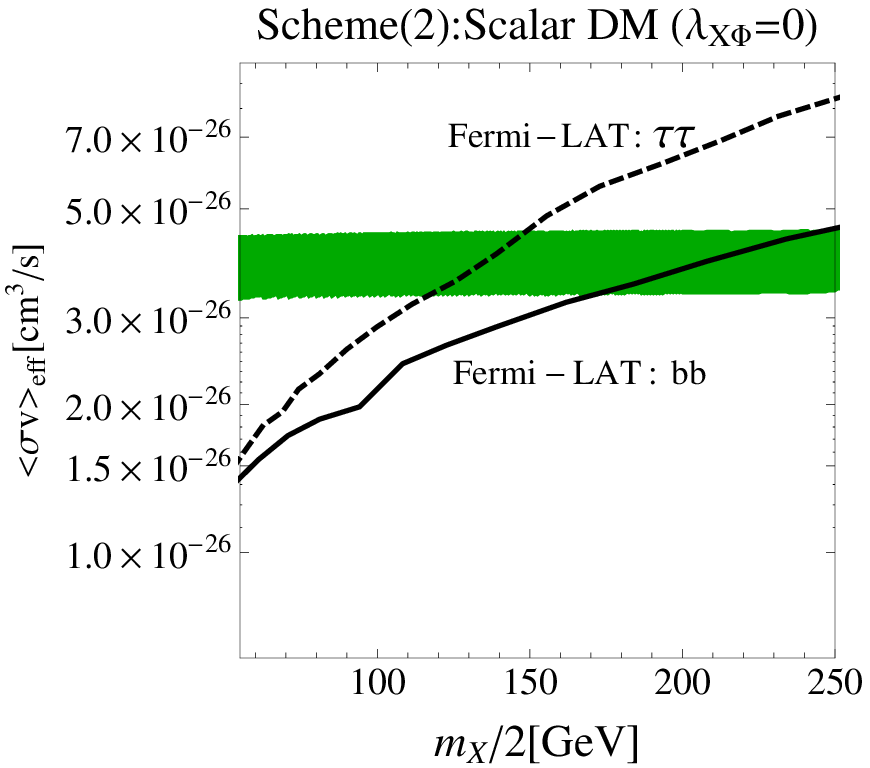} 
\includegraphics[width=80mm]{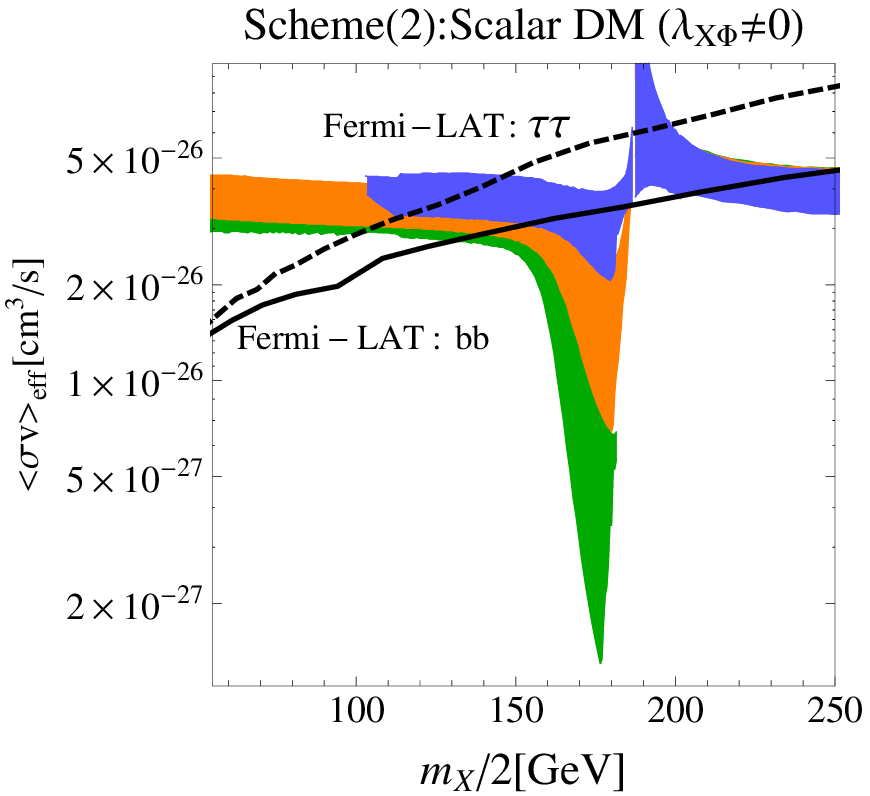} 
\caption{ The effective thermally averaged cross section for $\langle \sigma v \rangle_{\rm eff}$ of Eq.~(\ref{eq:sigma_eff}) in current Universe where the parameter region in Fig.~\ref{fig:RD1} is applied. The black solid(dashed) line indicate the current limit of the cross section for $DM DM \to b \bar b (\tau \bar \tau)$ annihilation mode from Fermi-LAT }
\label{fig:IDs2}
\end{center}
\end{figure}

For the scheme (2) we calculate the thermally averaged DM annihilation cross section $\langle \sigma v \rangle$ at the current Universe using the { \tt micrOMEGAs 4.1.5 } 
where we apply the parameters which is consistent with relic density of DM.
In this scheme the DM pair annihilate into $Z'$ pair dominantly, which provides 4 SM fermions in final states.
To discuss constraints from indirect detection experiments, we consider following effective cross section 
\begin{equation}
\label{eq:sigma_eff}
\langle \sigma v \rangle_{\rm eff} =   \langle \sigma v \rangle_{DM DM \to Z' Z'} 
\left[ 2 BR(Z'Z' \to 4 f^C_{SM}) + BR(Z' Z' \to 2 f^C_{SM} 2 f^N_{SM}) \right]
\end{equation}
where $BR(Z'Z' \to 4 f^C_{SM})$ and $BR(Z' Z' \to 2 f^C_{SM} 2 f^N_{SM})$ are branching fractions for both $Z'$ decaying into charged SM particles and 
that of one $Z'$ decaying into charged SM particles while one $Z'$ decaying into neutrinos, respectively, and the factor of 2 corresponds to doubled charged flux from $Z'$ decay. 
The Fig.~\ref{fig:IDs2} shows the $\langle \sigma v \rangle_{\rm eff}$ for the parameter region in Fig.~\ref{fig:RD1} which are compared with the constraints from Fermi-LAT for $DM DM \to b \bar b(\tau \bar \tau)$ annihilation mode~\cite{Ackermann:2015zua}; the constraints from light quark modes are similar to $b \bar b$ mode while that from electron and muon pair modes are weaker than the tau pair mode.
We note that our cross section can not be directly compared with the constraints from single annihilation mode since our $Z'$ decays all SM fermions.
Notice also that we compare the effective cross section at $m_X$ with experimental limits at $m_X/2$ since our final states have 4 particles and one particle carry energy of $m_X/2$.
Conservatively, we can say that the regions $m_X \lesssim 200$ GeV are disfavored. 
As in the scheme (1), the current thermally averaged annihilation cross section for fermion DM is much smaller than the constraint by Fermi-LAT, 
and we omit the plots for the case.

%%%%%%%%%%%%%%%%%%%%%%%%%%%%%%%%%%%%%%%%%%%%%%%%%%%%%%%%%%%%%%%%%%
%%%%%%%%%%%%%%%%%%%%%%%%%%%%%%%%%%%%%%%%%%%%%%%%%%%%%%%%%%
%%%%%%%%%%%%%%%%%%%%%%%%%%%%%%%%%%%%%%%%%%%%%%%%%%%%%%%%%%%%%%%%%%
%%%%%%%%%%%%%%%%%%%%%%%%%%%%%%%%%%%%%%%%%%%%%%%%%%%%%%%%%%

\section{The constraints from diphoton resonance search and implication to collider physics}
%%%%%%%%%%%%%%%%%%%%%%%%%%%%%%%%%%%%%%%%%%%%%%%%%%%%%%%%%%%%%%%%%%
\begin{figure}[tb] 
\begin{center}
\includegraphics[width=70mm]{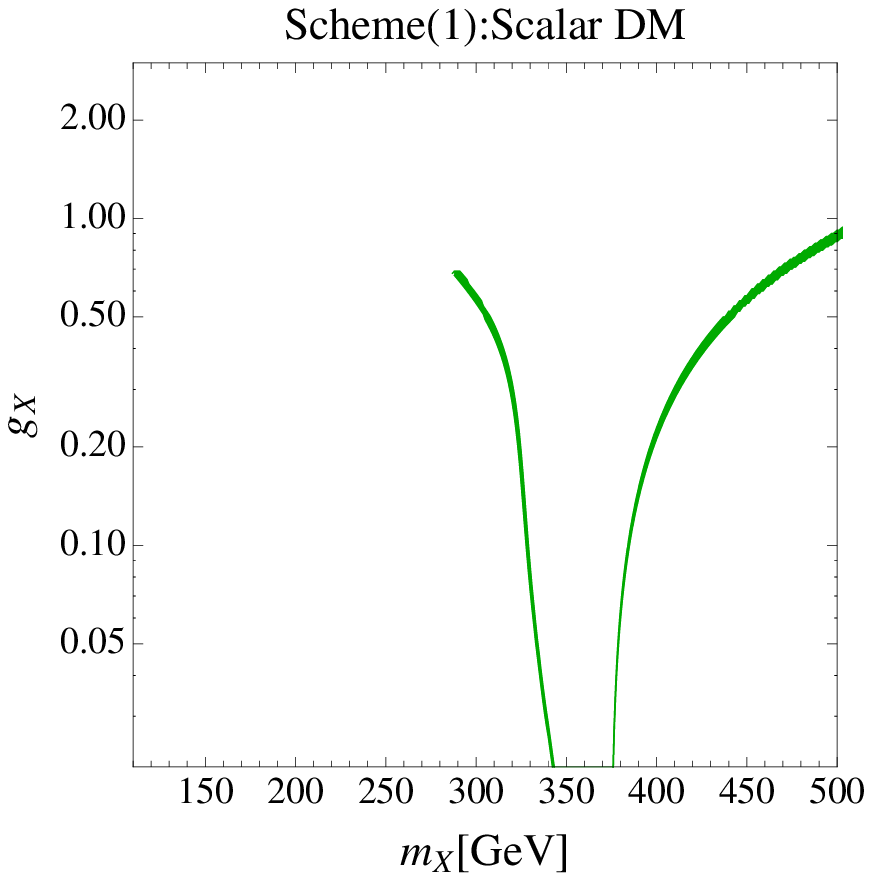} 
\includegraphics[width=70mm]{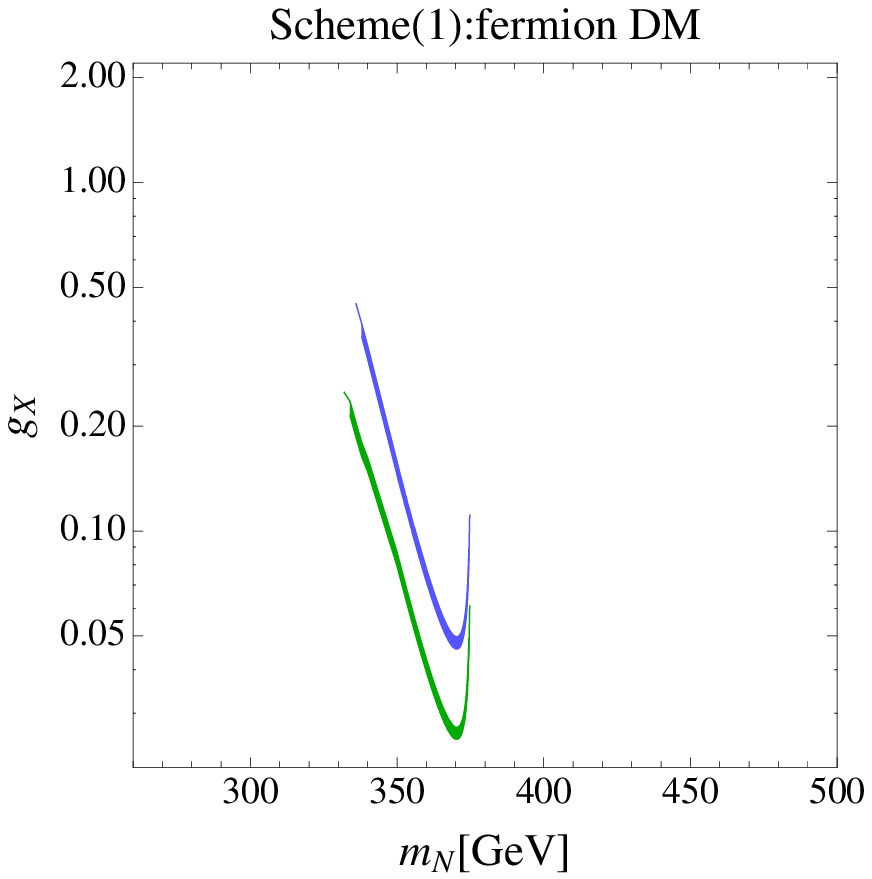} 
\caption{The parameter region which can accommodate relic density of DM and  constraints from collider experiment for scalar and fermion DM in scheme (1). }
\label{fig:Alloweds1}
\end{center}
\end{figure}
%%%%%%%%%%%%%%%%%%%%%%%%%%%%%%%%%%%%%%%%%%%%%%%%%%%%%%%%%%
%%%%%%%%%%%%%%%%%%%%%%%%%%%%%%%%%%%%%%%%%%%%%%%%%%%%%%%%%%%%%%%%%%
\begin{figure}[tb] 
\begin{center}
\includegraphics[width=70mm]{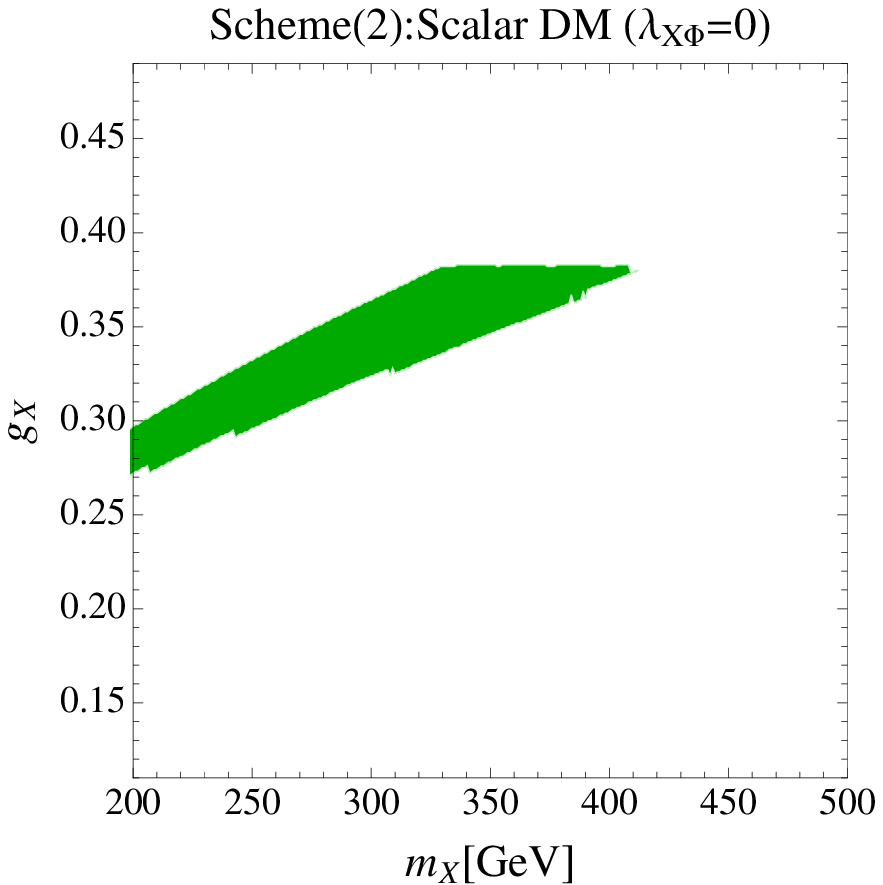} 
\includegraphics[width=70mm]{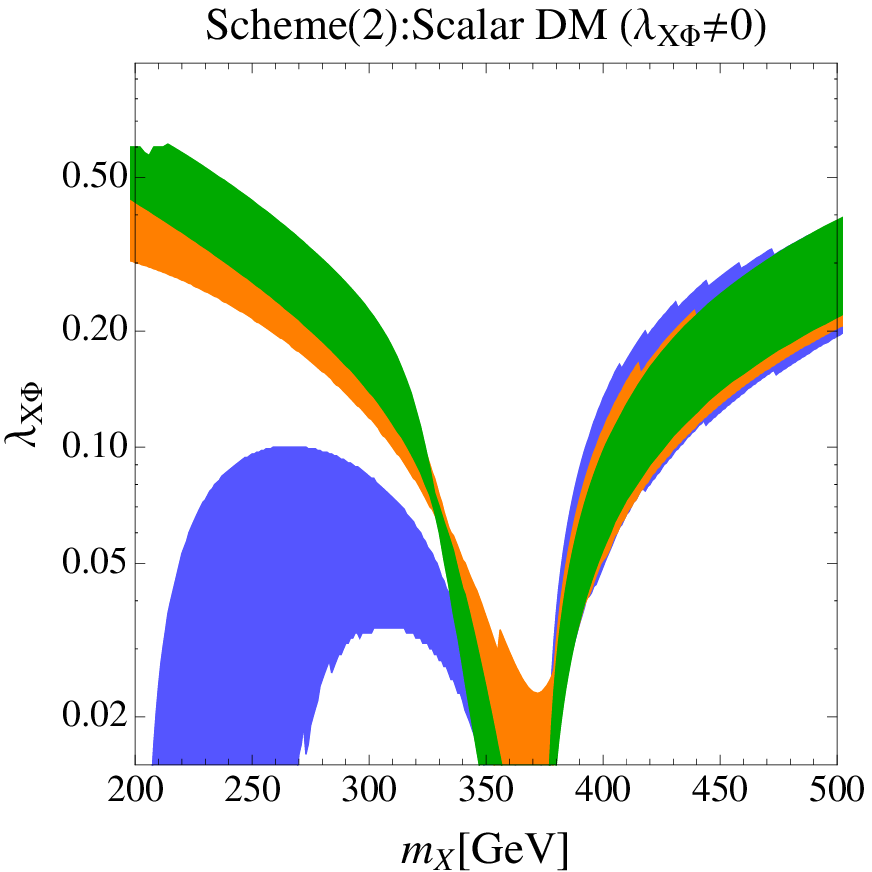} 
\caption{The parameter region which can accommodate thermal relic density of DM and  constraints from collider experiment 
for scalar DM in scheme (2). }
\label{fig:Alloweds2}
\end{center}
\end{figure}
%%%%%%%%%%%%%%%%%%%%%%%%%%%%%%%%%%%%%%%%%%%%%%%%%%%%%%%%%%
%%%%%%%%%%%%%%%%%%%%%%%%%%%%%%%%%%%%%%%%%%%%%%%%%%%%%%%%%%%%%%%%%%
\begin{figure}[tb] 
\begin{center}
\includegraphics[width=70mm]{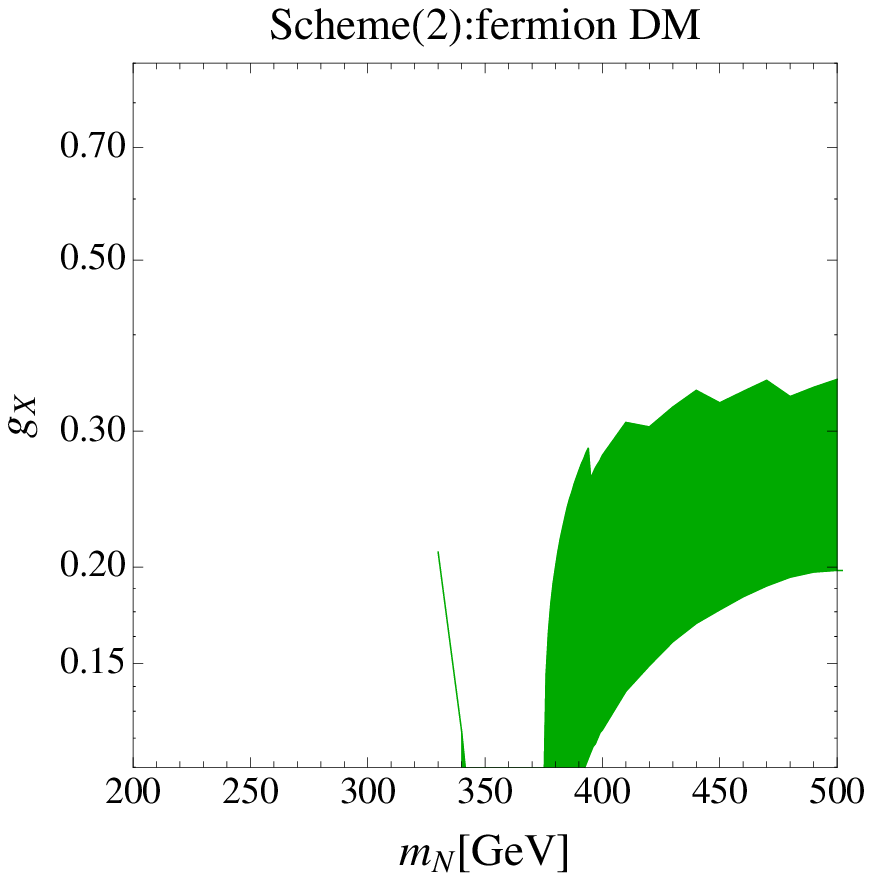} 
\caption{The parameter region which can accommodate relic density of DM and constraints from collider experiment for fermion DM in scheme (2). }
\label{fig:Alloweds3}
\end{center}
\end{figure}
%%%%%%%%%%%%%%%%%%%%%%%%%%%%%%%%%%%%%%%%%%%%%%%%%%%%%%%%%%
%
In this section we discuss the constraint from current data of diphoton resonance search and search for the parameter region 
which is consistent with both the diphoton data and DM physics.  
Then we shall consider the collider signatures in terms of $\phi$ production cross section in that parameter 
region.

\subsection{The constraint from diphoton resonance search}

Here we discuss the constraints from diphoton resonance search in the model and search for the parameter region which is consistent 
with constraints from DM physics.  In our scenario, the scalar boson $\phi$ provides diphoton resonance where mass of $\phi$ is set to 750 GeV as we mentioned above.  
$\phi$ can be produced by gluon fusion process through effective interaction 
Eq.~(\ref{eq:LggS}) at the LHC.
The decay mode of $\phi \to \gamma \gamma$ is induced by the new fermion loop same as the $\phi \to gg$.
Then we obtain the decay width of the diphoton mode as Eq.~(\ref{eq:width-diphoton}).
In the narrow width approximation, the cross section for the process $pp \to \phi \to \gamma \gamma$ 
through gluon fusion can be expressed as~\cite{Franceschini:2015kwy} 
\begin{equation}
\label{eq:CXphi}
\sigma(pp \to \phi \to \gamma \gamma) \simeq \frac{C_{gg}}{s} \frac{\Gamma_{\phi \to gg}}{m_\phi} BR(\phi \to \gamma \gamma)
\end{equation}
where $C_{gg}$ is related to the gluon luminosity function, $s$ is the center of energy and $BR(\phi \to \gamma \gamma)$ is the branching fraction of $\phi \to \gamma \gamma$ decay.
For $\sqrt{s} = 13(8)$ TeV, we adopt $C_{gg} \simeq 2137(174)$.
In addition, we apply the K-factor for gluon fusion process as $K_{gg} \simeq 1.5$~\cite{Franceschini:2015kwy}. 
Here we search for the parameter region which is allowed by the current data of the diphoton resonance search and consistent with constraints from DM physics.
We then estimate $\sigma (gg\to \phi \to \gamma \gamma)$ applying parameter space which is consistent with constraints from DM physics for both schemes (1) and (2)
in order to search for the region which is allowed by the current constraint from diphoton resonance search.
To satisfy the constraint, we require the cross section to be 
\begin{equation}
\label{eq:CXdiphoton}
\sigma(gg \to \phi \to \gamma \gamma) \leq 1.21 \, {\rm fb},
\end{equation}
where we take into account 1$\sigma$ error of ATLAS result in Ref.~\cite{ATLAS:2016eeo}.
We also applied constraints on the cross section for $pp \to \phi \to {\rm invisible}$ from the mono-jet search data at the LHC 8 TeV~\cite{Khachatryan:2014rra}:
\begin{equation}
\label{eq:const_invisible}
\sigma (pp \to \phi \to {\rm invisible}) < 0.8 \, {\rm pb}.
\end{equation}
We note that the process $pp \to \phi \to gg$ provides dijet final state but the cross section in our model are smaller than constraint from current dijet search at the center-of-energy of $\sqrt{s}=8$ and 13 TeV ~\cite{Aad:2014aqa, Khachatryan:2015dcf,ATLAS:2015nsi}.
Here we comment on the case where $\phi$ is rather heavy, e.g., $m_{\phi} = 1.0$ 
and $1.5$ TeV. In these cases, $\phi$ production cross section becomes $\sigma_{m_{\phi} 
=1.0(1.5) \, {\rm TeV}} \simeq 0.42(0.19) \times \sigma_{m_\phi = 750 \, {\rm GeV}}$ when other 
parameters are taken to be the same values as before. On the other hand, current upper limit for 
$\sigma(gg \to \phi \to \gamma \gamma)$ by ATLAS is $0.78(0.48)$ fb for $m_\phi = 1.0(1.5)$ TeV. 
Therefore the constrains from diphoton mode is weaker for heavier $m_\phi$ since production 
cross section  rapidly decreases compared with the change of the upper limit.  

In the scheme (1), we obtain the allowed region shown in Figs.~\ref{fig:Alloweds1} for scalar and fermionic 
DM  cases, which is consistent with the diphoton constraint and DM physics.
%The  lighter colored regions satisfy only the upper limit by diphoton channel of $\phi \to \gamma \gamma$ 
%as $\sigma(gg \to \phi \to \gamma \gamma) < 7$ fb.
We find that the region $m_X \lesssim 300$ GeV is excluded by Eq.~(\ref{eq:const_invisible}) for scalar DM 
case.   Moreover most of the parameter region is excluded by the diphoton constraint except for the region of 330 GeV $ \lesssim m_N \lesssim 380$ GeV for the fermionic DM case.  
We note that in this scheme the width of $\phi$ is narrower than $1$ GeV since $\phi$ decays via loop effects.

In the scheme (2), we obtain the allowed region in Fig.~\ref{fig:Alloweds2} and \ref{fig:Alloweds3} for scalar 
and fermion DM cases respectively.
For scalar DM case with $\lambda_{X\Phi} =0$, the region of $m_X \gtrsim 420$ GeV is excluded by the diphoton constraint.
For scalar DM case with $\lambda_{X \Phi} \neq 0$, we find that the DM mass region 200 GeV $\geq m_X \geq 500$ GeV can be accommodated with the constraints Eqs~(\ref{eq:CXdiphoton}) and (\ref{eq:const_invisible}).
For the fermion DM case, we find that the region $m_X \lesssim 340$ GeV is excluded by Eq.~(\ref{eq:const_invisible}).
We note that in this scheme the width of $\phi$ is $O(10)$ to $O(50)$ GeV for $m_\phi > 2 m_{Z'}$ 
and less than 1 GeV for $m_\phi < 2 m_{Z'}$.

\subsection{The $\phi$ production cross section}
%%%%%%%%%%%%%%%%%%%%%%%%%%%%%%%%%%%%%%%%%%%%%%%%%%%%%%%%%%%%%%%%%%
\begin{figure}[tb] 
\begin{center}
\includegraphics[width=70mm]{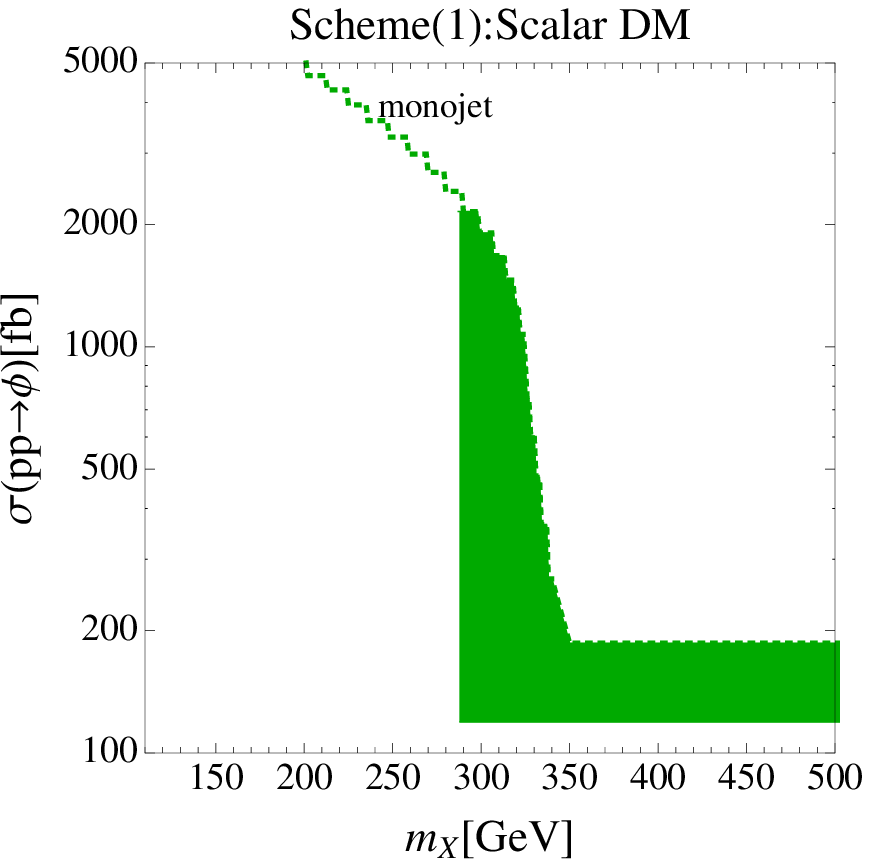} 
\includegraphics[width=70mm]{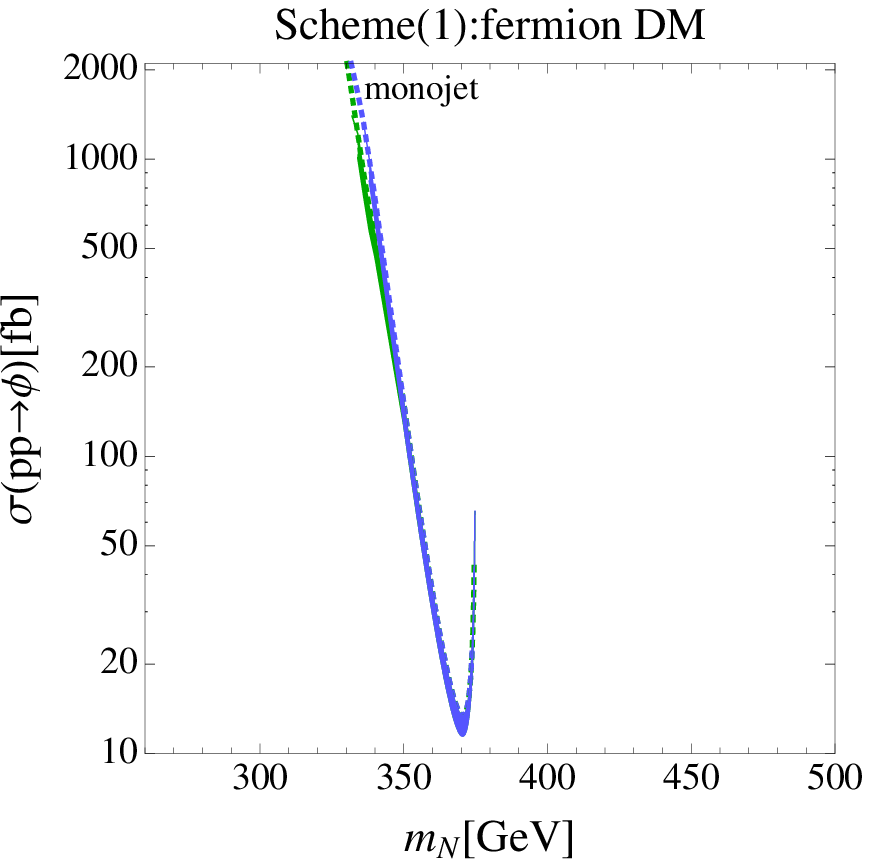} 
\caption{The $\phi$ production cross section for the parameter region in Fig.~\ref{fig:Alloweds1}. 
The dotted lines indicate the upper limits on the cross section from 
$pp \to \phi \to XX$ channel with mono-jet search data at the LHC~\cite{Khachatryan:2014rra}. }
\label{fig:CXphiS1}
\end{center}
\end{figure}
%%%%%%%%%%%%%%%%%%%%%%%%%%%%%%%%%%%%%%%%%%%%%%%%%%%%%%%%%%
%%%%%%%%%%%%%%%%%%%%%%%%%%%%%%%%%%%%%%%%%%%%%%%%%%%%%%%%%%%%%%%%%%
\begin{figure}[tb] 
\begin{center}
\includegraphics[width=70mm]{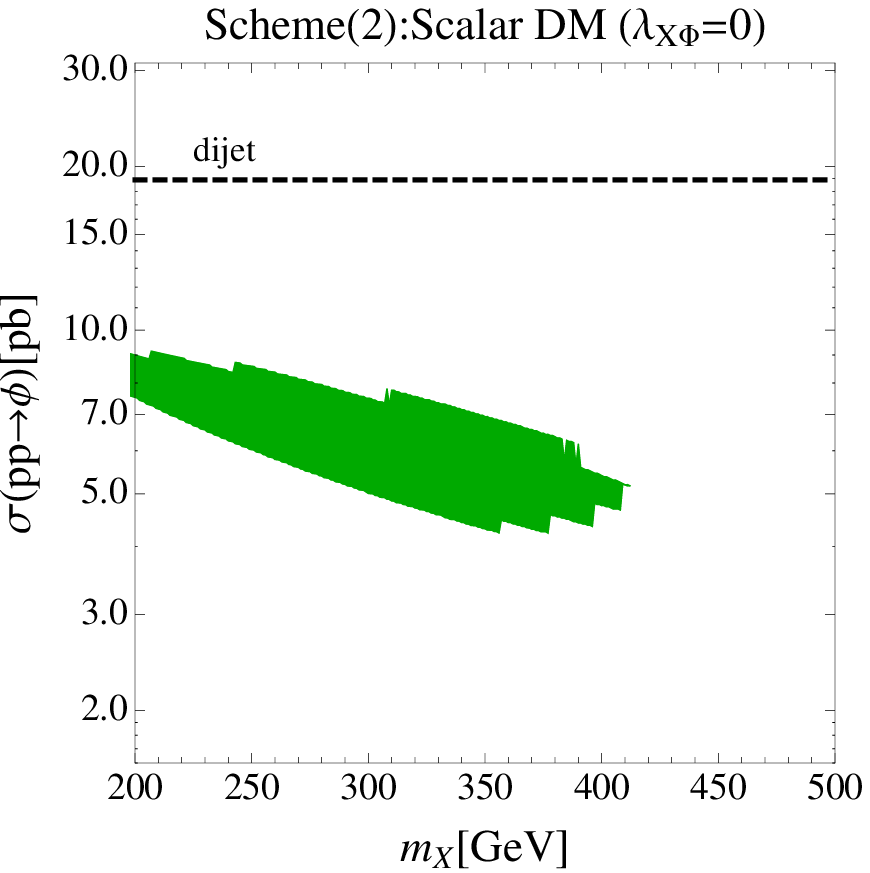} 
\includegraphics[width=70mm]{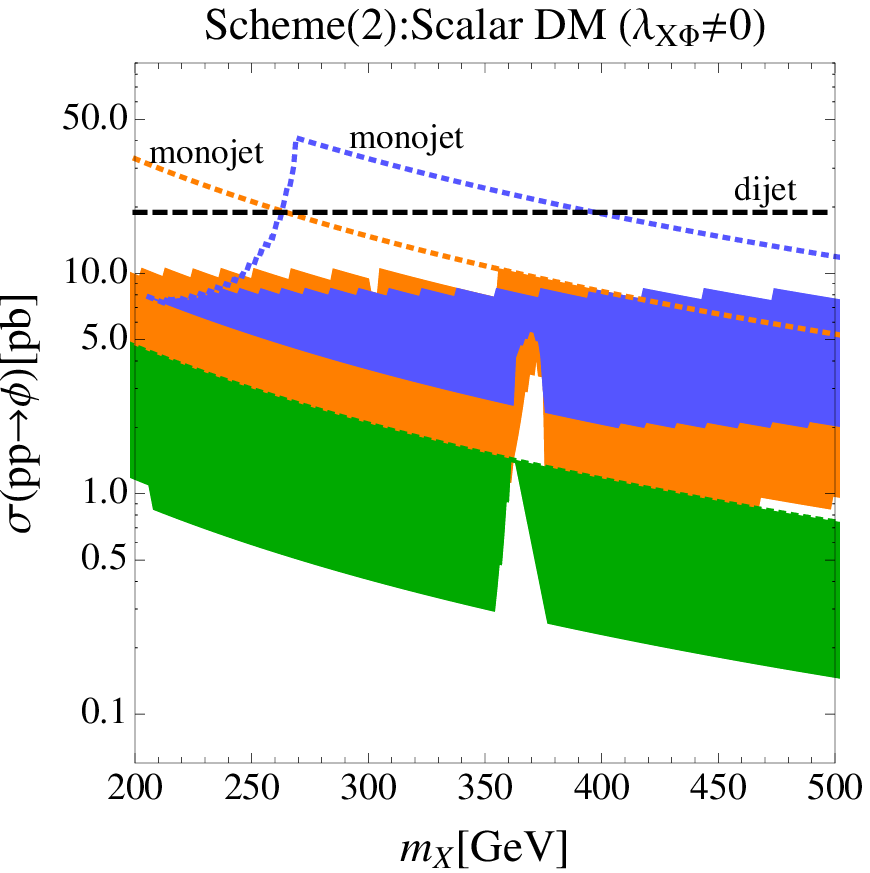} 
\caption{The $\phi$ production cross section for the parameter region in Fig.~\ref{fig:Alloweds2}. The dashed line indicates the upper limit on the cross section from dijet search 
at the LHC~\cite{Aad:2014aqa} for $BR(\phi \to jj)=1.0$. The dotted lines for right figure are 
the mono-jet constraint as in Fig.~\ref{fig:CXphiS1}.}
\label{fig:CXphiS2}
\end{center}
\end{figure}
%%%%%%%%%%%%%%%%%%%%%%%%%%%%%%%%%%%%%%%%%%%%%%%%%%%%%%%%%%
%%%%%%%%%%%%%%%%%%%%%%%%%%%%%%%%%%%%%%%%%%%%%%%%%%%%%%%%%%%%%%%%%%
\begin{figure}[tb] 
\begin{center}
\includegraphics[width=70mm]{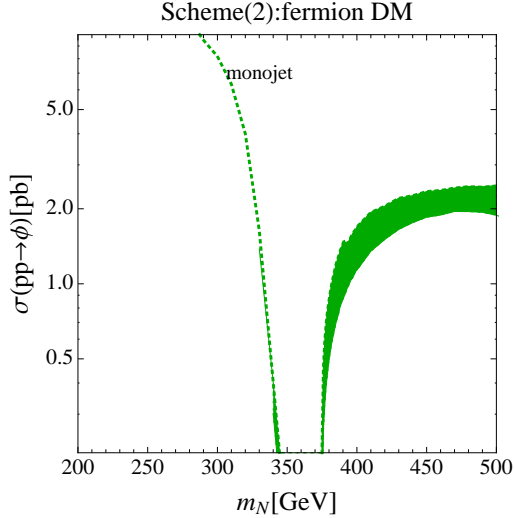} 
\caption{The $\phi$ production cross section for the parameter region in Fig.~\ref{fig:Alloweds3}.
The dotted line indicates  the mono-jet constraints as in Fig.~\ref{fig:CXphiS1} and 
in \ref{fig:CXphiS2}.}
\label{fig:CXphiS2N}
\end{center}
\end{figure}
%%%%%%%%%%%%%%%%%%%%%%%%%%%%%%%%%%%%%%%%%%%%%%%%%%%%%%%%%%
Here we explore the $\phi$ production cross section for the parameter region which is consistent with the 
constraints from collider experiment and DM physics. 
The $\phi$ production cross section is derived from Eq.~(\ref{eq:CXphi}).
For the scheme (1), we show the cross sections in Fig.~\ref{fig:CXphiS1} which is obtained by applying the parameter space shown in Fig.~\ref{fig:Alloweds1}. 
We then have $O(100)(O(10))$ fb to $O(1000)$ fb cross section for scalar (fermion) DM cases for the allowed region.
Since $\phi$ mainly decays into two gluons in the scheme (1), the dijet event is the another signature of $\phi$ as $pp \to \phi \to jj$ which can be tested at the LHC.

For the scheme (2), we obtain the cross sections shown in Figs.~\ref{fig:CXphiS2} and \ref{fig:CXphiS2N} 
which are obtained by applying the parameter spaces in Figs.~\ref{fig:Alloweds2} and \ref{fig:Alloweds3}, 
respectively.    
We then find that $\sim 4$ pb to $\sim 9$ pb cross section is obtained for the allowed region 
for $\lambda_{X\Phi}=0$,  while $\sim 0.2$ pb to $\sim 10$ pb cross section is obtained for $\lambda_{X\Phi} \neq 0$.
These cross sections would be constrained when $\phi \to Z' Z'$ mode is kinematically allowed, since $Z'$ 
can decay into SM leptons with $BR(\phi \to \ell^+ \ell^-)\sim O(20) \%$  via kinetic mixing~\cite{Ko:2016wce} 
inducing signal events such as  $2j + ll, 4 l, 2l + \ET$. We also have signal events $4j, 2j + \ET$ but it will be 
less significant due to large SM backgrounds.
The detailed analysis of current experimental constraints for $pp \to \phi \to Z' Z'$ process is beyond the scope 
of this paper and we left it as future work.   We also would like to comment that the branching fraction of $Z'$ 
would be modified with non-zero Yukawa coupling $F f_{SM} X$ at loop level since the kinetic mixing is very small.

\section{Summary and discussion}

In this paper, we have investigated dark matter physics for the chiral dark sector model where dark fermions are vectorlike under SU(3)$\times$U(1)$_Y$ but chiral under dark U(1)$_X$ gauge symmetry.
In our model, the extra scalar bosons with non-zero U(1)$_X$ ($\Phi$ and $X$ in Table I) are also introduced 
in order to break the U(1) symmetry spontaneously and to make dark fermions decay respectively. 
As a result of our set up, we have accidental $Z_2$ symmetry which guarantees the stability of our DM 
candidate: scalar boson $X$ and neutral dark fermion $N$.
We also have massive new gauge boson $Z'$ after spontaneous breaking of U(1)$_X$ gauge symmetry.

In our analysis of DM physics, two different schemes are considered: (1) $m_{DM} < m_{Z'}$ and 
(2) $m_{DM} > m_{Z'}$, where $m_{DM}$ and $m_{Z'}$ are DM mass and mass of $Z'$ boson respectively.
For the scheme (1), the dominant DM annihilation process is $DM DM \to g g$ exchanging scalar boson 
$\phi$ associated with U(1)$_X$ breaking, while the dominant annihilation process for the scheme (2) is 
$DM DM \to Z' Z'$. 
Then we have investigated the relic density of DM, DM-nucleon scattering cross section for direct detection 
and  DM annihilation cross section for indirect detection searching for the parameter region 
which is allowed by constraints from current observations.  
In our analysis we fixed dark fermion masses to reduce the number of free parameters, and explored the 
parameter space of DM masses, $Z'$ mass, $g_X$ and coupling constant $\lambda_{X\Phi}$ for $\Phi \Phi X X$ 
interaction  in both schemes.

For  the scheme (1), the relic density of DM is determined by the annihilation cross section for 
$XX(NN) \to gg$ processes.   Then we have shown allowed parameter space providing observed relic 
density in  the $m_X$-$\lambda_{X\Phi}$ and $m_N$-$g_X$ planes for scalar and fermion DM cases respectively.  
We find that $m_{X} \gtrsim 140$ GeV region can provide the observed relic density with $\lambda_{X\Phi} 
< \sqrt{4 \pi}$ for scalar DM, while fermion DM can have the right relic density in all DM mass region of 
our consideration. These parameter region are further constrained by direct and indirect detection experiment. 
The parameter region with $m_{X} \lesssim 200$ GeV is excluded for scalar DM, whereas $m_{N} \lesssim 300$ 
GeV region is excluded for fermion DM when $m_{Z'} =1.1 m_N$.
On the other hand, constraints from indirect detection exclude region with $m_X \simeq m_\phi/2$ for scalar 
DM, whereas no further constraint is imposed for fermion DM.

For the scheme (2), thermal relic density is determined by the annihilation cross section for $XX(NN) \to Z' Z'$ 
processes.   Then the allowed parameter regions giving right relic density for scalar DM are shown in the 
$m_X$-$g_X$ plane (with $\lambda_{X\Phi}=0$) and $m_X$-$\lambda_{X\Phi}$ (with several values of $g_X$) 
while the allowed parameter region for fermion DM is shown in the $m_N$-$g_X$ plane. 
Then we have shown that the current constraints from direct detection exclude some region with 
$m_X \lesssim 150$ GeV and $g_X =0.1$ for scalar DM,   and the DM-nucleon scattering cross section is 
below current limit for fermion DM.
We also find that the parameter region with $m_{X} \lesssim 200$ GeV is constrained by the Fermi-LAT data 
for scalar DM while fermion DM is free from indirect detection constraints due to absence of a $s$-channel annihilation mode.   The parameter spaces can be further tested in future direct and indirect detection 
experiments for both schemes.

Finally we have also investigated constraints from the collider experiment including diphton resonance search by ATLAS and CMS at the LHC 
13 TeV and searched for the parameter region which can accommodate both DM and collider constraints.   
In our model a source of diphoton resonance is scalar boson $\phi$, which is nothing 
but a remnant of U(1)$_X$ breaking by dark Higgs mechanism.  
The cross section for $pp \to \phi \to \gamma \gamma$ is estimated applying parameter sets that can provide 
the correct thermal relic density of DM.  Then we have shown the parameter regions which can accommodate with 
the constraint from diphoton resonance search for both schemes  with scalar and fermion DM.  
In addition, we have discussed $\phi$ production cross section applying the parameter regions.
We find that the cross section in the scheme (2) tends to be larger than that in the scheme (1).  The signatures 
of $\phi$ other than diphoton event are dijet and $Z'Z'$ events for the scheme (1) and the scheme (2) 
respectively, where $Z'$ decays into SM fermion pair thereby providing 4 SM fermion final states.
Detailed analysis of signals and backgrounds is beyond the scope of this paper and we left it as future works.  

Before closing, we comment on the stability of the potential in our model. In the  previous study~\cite{Ko:2016wce}, we discussed the stability of the scalar potential within the  renormalization group running and found our model could be valid up to $\sim O(10)$ TeV with 
$O(1)$ Yukawa couplings and other couplings in the scalar potential.  In the present analysis, 
we can take smaller couplings since the cross section for $pp \to \phi \to \gamma \gamma$ 
should be smaller than our previous analysis, and so the stability could be achieved upto higher 
scale.

\section*{Acknowledgments} 
This work is supported in part by National Research Foundation of Korea (NRF) Research Grant NRF-2015R1A2A1A05001869, and by the NRF grant funded by the Korea government (MSIP) 
(No. 2009-0083526) through Korea Neutrino Research Center at Seoul National University (PK).

\appendix

\renewcommand{\theequation}{A.\arabic{equation} }
\setcounter{equation}{0}

\section*{Appendix: The decay widths of $\phi$}

Here we summarize the decay widths of $\phi$ which are given in~\cite{Ko:2016wce}.
The width for $\phi \to gg$ mode is given by 
\begin{equation}
\label{eq:width-gg}
\Gamma_{\phi \rightarrow gg}  = \frac{ \alpha_s^2 m_\phi^3 }{32 \pi^3}  \left|\sum_{F=U,D}\frac{ (a+b) g_X}{2 m_{Z'}}    A_{1/2}(\tau_F) \right|^2.
  \end{equation}  
Similarly the partial decay width for $\phi \to \gamma \gamma$ is given by
\begin{equation}
\label{eq:width-diphoton}
\Gamma_{\phi \to \gamma\gamma} = \frac{\alpha^2 m^3_\phi}{256 \pi^3} \left| \sum_F N_c^F \frac{(a+b)g_X Q^2_{F}}{m_{Z'}} A_{1/2}(\tau_F) \right|^2 \,,
  \end{equation}  
  where $Q_F$ and $N_c^F$ are electric charge and number of color of an exotic fermions $F$. 
The formula for the partial decay width of $\phi \to Z\gamma$ is 
\begin{align}
 \Gamma_{\phi \to Z\gamma} =& \frac{ m^3_\phi }{32\pi} \left| A_{Z\gamma}\right|^2  
\left(1 - \frac{m_Z^2}{m_\phi^2}\right)^3 \,, \\
 A_{Z\gamma} =& \frac{2\sqrt{2} \alpha s_W g_X}{\pi c_W}  \sum_{F} \frac{N_c^F (a+b) Q_{F}^2}{m_{Z'}} [I_1 ( \tau_{F}, \lambda_{F}) -I_2(\tau_{F}, \lambda_{F}) ]\,, \nonumber
\end{align}
where $\lambda_{F}= 4 m^2_{F}/m^2_Z$ and the corresponding loop integrals are given by~\cite{Gunion:1989we}:
 \begin{align}
 I_1(x,y) =& \frac{xy}{2(x-y)} + \frac{x^2 y^2}{2(x-y)^2} [ f(x)^2 - f(y)^2 ] + \frac{x^2 b}{(x-y)^2} [g(x)-g(y)]\,,\nonumber \\
 I_2(x,y) =& - \frac{x y}{2(x-y)} [ f(x)^2 - f(y)^2 ] \,, \nonumber \\
 g(t) =& \sqrt{t -1} \sin^{-1}(1/\sqrt{t})\,. 
 \end{align}
The decay widths of $\phi$ into $Z'Z'$, $X^*X$ and $\bar F F$ modes are obtained 
at tree level such that 
\begin{align}
\Gamma_{\phi \to Z' Z'} =& \frac{(a+b)^2 g_X^2 m_{Z'}^2}{32 \pi m_\phi} \frac{m_\phi^4 - 4 m_\phi^2 m_{Z'}^2 + 12 m_{Z'}^4}{m_{Z'}^4} \sqrt{1 - \frac{4 m_{Z'}^2}{m_\phi^2}}\,, \\
\Gamma_{\phi \to X^* X} =& \frac{\lambda_{X\Phi}^2 m_{Z'}^2}{16 \pi (a+b)^2 g_X^2 m_\phi} 
\sqrt{1 - \frac{4 m_X^2}{m_\phi^2}}, \\
\Gamma_{\phi \to \bar F F} =& \frac{g_X^2 M_F^2}{4 \pi m_{Z'}^2} m_\phi \sqrt{1 - \frac{4 M_F^2}{m_{Z'}^2}}.
\end{align}


\begin{thebibliography}{99}

  %\cite{Aaboud:2016tru}
\bibitem{Aaboud:2016tru} 
  M.~Aaboud {\it et al.} [ATLAS Collaboration],
  %``Search for resonances in diphoton events at $\sqrt{s}$=13 TeV with the ATLAS detector,''
  arXiv:1606.03833 [hep-ex].
  %%CITATION = ARXIV:1606.03833;%%
  %19 citations counted in INSPIRE as of 13 Jul 2016
  
  %\cite{Khachatryan:2016hje}
\bibitem{Khachatryan:2016hje} 
  V.~Khachatryan {\it et al.} [CMS Collaboration],
  %``Search for resonant production of high-mass photon pairs in proton-proton collisions at sqrt(s) = 8 and 13 TeV,''
  arXiv:1606.04093 [hep-ex].
  %%CITATION = ARXIV:1606.04093;%%
  %21 citations counted in INSPIRE as of 13 Jul 2016
  
  %\cite{}
\bibitem{ATLAS:2016} 
  The ATLAS collaboration,
  %``Search for resonances in diphoton events with the ATLAS detector at $\sqrt{s}$ = 13 TeV,''
  ATLAS-CONF-2016-018.
  %%CITATION = ATLAS-CONF-2016-018;%%
  %4 citations counted in INSPIRE as of 11 Apr 2016
  
  %\cite{CMS:2016owr}
\bibitem{CMS:2016owr} 
  CMS Collaboration [CMS Collaboration],
  %``Search for new physics in high mass diphoton events in $3.3~\mathrm{fb}^{-1}$ of proton-proton collisions at $\sqrt{s}=13~\mathrm{TeV}$ and combined interpretation of searches at $8~\mathrm{TeV}$ and $13~\mathrm{TeV}$,''
  CMS-PAS-EXO-16-018.
  %%CITATION = CMS-PAS-EXO-16-018;%%
  %11 citations counted in INSPIRE as of 11 Apr 2016
  

%%%%%%%%%%%%%%%%%%%%%%%%%%%%%%%%%%%%%%%%%%%%%%%%%%%%%%


  %\cite{Ko:2016wce}
\bibitem{Ko:2016wce} 
  P.~Ko and T.~Nomura,
  %``Dark sector shining through 750 GeV dark Higgs boson at the LHC,''
  Phys.\ Lett.\ B {\bf 758}, 205 (2016)
%  doi:10.1016/j.physletb.2016.05.014
  [arXiv:1601.02490 [hep-ph]].
  %%CITATION = doi:10.1016/j.physletb.2016.05.014;%%
  %44 citations counted in INSPIRE as of 15 Jun 2016


 %%%%%%%%%%%%%%%%%%%%%%%%%%%%%%%%%%%%%%%%%%%%%%%%%%%%
 
 
    %\cite{Franceschini:2015kwy}
\bibitem{Franceschini:2015kwy} 
  R.~Franceschini {\it et al.},
  %``What is the $\gamma \gamma$ resonance at 750 GeV?,''
  JHEP {\bf 1603}, 144 (2016)
%  doi:10.1007/JHEP03(2016)144
  [arXiv:1512.04933 [hep-ph]].
  %%CITATION = doi:10.1007/JHEP03(2016)144;%%
  %279 citations counted in INSPIRE as of 21 May 2016
 
 
 %\cite{Harigaya:2015ezk}
\bibitem{Harigaya:2015ezk} 
  K.~Harigaya and Y.~Nomura,
  %``Composite Models for the 750 GeV Diphoton Excess,''
  Phys.\ Lett.\ B {\bf 754}, 151 (2016)
%  doi:10.1016/j.physletb.2016.01.026
  [arXiv:1512.04850 [hep-ph]].
  %%CITATION = doi:10.1016/j.physletb.2016.01.026;%%
  %208 citations counted in INSPIRE as of 12 Jun 2016
  
  %\cite{Backovic:2015fnp}
\bibitem{Backovic:2015fnp} 
  M.~Backovic, A.~Mariotti and D.~Redigolo,
  %``Di-photon excess illuminates Dark Matter,''
  JHEP {\bf 1603}, 157 (2016)
%  doi:10.1007/JHEP03(2016)157
  [arXiv:1512.04917 [hep-ph]].
  %%CITATION = doi:10.1007/JHEP03(2016)157;%%
  %188 citations counted in INSPIRE as of 12 Jun 2016
  
  %\cite{Angelescu:2015uiz}
\bibitem{Angelescu:2015uiz} 
  A.~Angelescu, A.~Djouadi and G.~Moreau,
  %``Scenarii for interpretations of the LHC diphoton excess: two Higgs doublets and vector-like quarks and leptons,''
  Phys.\ Lett.\ B {\bf 756}, 126 (2016)
%  doi:10.1016/j.physletb.2016.02.064
  [arXiv:1512.04921 [hep-ph]].
  %%CITATION = doi:10.1016/j.physletb.2016.02.064;%%
  %198 citations counted in INSPIRE as of 21 Jun 2016

 
 %\cite{Nakai:2015ptz}
\bibitem{Nakai:2015ptz} 
  Y.~Nakai, R.~Sato and K.~Tobioka,
  %``Footprints of New Strong Dynamics via Anomaly and the 750 GeV Diphoton,''
  Phys.\ Rev.\ Lett.\  {\bf 116}, no. 15, 151802 (2016)
%  doi:10.1103/PhysRevLett.116.151802
  [arXiv:1512.04924 [hep-ph]].
  %%CITATION = doi:10.1103/PhysRevLett.116.151802;%%
  %181 citations counted in INSPIRE as of 12 Jun 2016
  
  %\cite{Buttazzo:2015txu}
\bibitem{Buttazzo:2015txu} 
  D.~Buttazzo, A.~Greljo and D.~Marzocca,
  %``Knocking on new physics’ door with a scalar resonance,''
  Eur.\ Phys.\ J.\ C {\bf 76}, no. 3, 116 (2016)
%  doi:10.1140/epjc/s10052-016-3970-7
  [arXiv:1512.04929 [hep-ph]].
  %%CITATION = doi:10.1140/epjc/s10052-016-3970-7;%%
  %199 citations counted in INSPIRE as of 12 Jun 2016
  
  %\cite{DiChiara:2015vdm}
\bibitem{DiChiara:2015vdm} 
  S.~Di Chiara, L.~Marzola and M.~Raidal,
  %``First interpretation of the 750 GeV diphoton resonance at the LHC,''
  Phys.\ Rev.\ D {\bf 93}, no. 9, 095018 (2016)
%  doi:10.1103/PhysRevD.93.095018
  [arXiv:1512.04939 [hep-ph]].
  %%CITATION = doi:10.1103/PhysRevD.93.095018;%%
  %186 citations counted in INSPIRE as of 12 Jun 2016
  
  %\cite{Knapen:2015dap}
\bibitem{Knapen:2015dap} 
  S.~Knapen, T.~Melia, M.~Papucci and K.~Zurek,
  %``Rays of light from the LHC,''
  Phys.\ Rev.\ D {\bf 93}, no. 7, 075020 (2016)
%  doi:10.1103/PhysRevD.93.075020
  [arXiv:1512.04928 [hep-ph]].
  %%CITATION = doi:10.1103/PhysRevD.93.075020;%%
  %207 citations counted in INSPIRE as of 12 Jun 2016
  
  %\cite{Pilaftsis:2015ycr}
\bibitem{Pilaftsis:2015ycr} 
  A.~Pilaftsis,
  %``Diphoton Signatures from Heavy Axion Decays at the CERN Large Hadron Collider,''
  Phys.\ Rev.\ D {\bf 93}, no. 1, 015017 (2016)
%  doi:10.1103/PhysRevD.93.015017
  [arXiv:1512.04931 [hep-ph]].
  %%CITATION = doi:10.1103/PhysRevD.93.015017;%%
  %174 citations counted in INSPIRE as of 12 Jun 2016
   
  %\cite{Ellis:2015oso}
\bibitem{Ellis:2015oso} 
  J.~Ellis, S.~A.~R.~Ellis, J.~Quevillon, V.~Sanz and T.~You,
  %``On the Interpretation of a Possible $\sim 750$ GeV Particle Decaying into $\gamma \gamma$,''
  JHEP {\bf 1603}, 176 (2016)
%  doi:10.1007/JHEP03(2016)176
  [arXiv:1512.05327 [hep-ph]].
  %%CITATION = doi:10.1007/JHEP03(2016)176;%%
  %213 citations counted in INSPIRE as of 12 Jun 2016
  
  %\cite{Gupta:2015zzs}
\bibitem{Gupta:2015zzs} 
  R.~S.~Gupta, S.~Jager, Y.~Kats, G.~Perez and E.~Stamou,
  %``Interpreting a 750 GeV Diphoton Resonance,''
  arXiv:1512.05332 [hep-ph].
  %%CITATION = ARXIV:1512.05332;%%
  %197 citations counted in INSPIRE as of 12 Jun 2016
  
  %\cite{Kobakhidze:2015ldh}
\bibitem{Kobakhidze:2015ldh}
  A.~Kobakhidze, F.~Wang, L.~Wu, J.~M.~Yang and M.~Zhang,
  %``750 GeV diphoton resonance in a top and bottom seesaw model,''
  Phys.\ Lett.\ B {\bf 757} (2016) 92
%  doi:10.1016/j.physletb.2016.03.067
  [arXiv:1512.05585 [hep-ph]].
  %%CITATION = doi:10.1016/j.physletb.2016.03.067;%%
  %163 citations counted in INSPIRE as of 16 Jun 2016
  
  %\cite{Falkowski:2015swt}
\bibitem{Falkowski:2015swt} 
  A.~Falkowski, O.~Slone and T.~Volansky,
  %``Phenomenology of a 750 GeV Singlet,''
  JHEP {\bf 1602}, 152 (2016)
%  doi:10.1007/JHEP02(2016)152
  [arXiv:1512.05777 [hep-ph]].
  %%CITATION = doi:10.1007/JHEP02(2016)152;%%
  %207 citations counted in INSPIRE as of 12 Jun 2016
  
  %\cite{Benbrik:2015fyz}
\bibitem{Benbrik:2015fyz} 
  R.~Benbrik, C.~H.~Chen and T.~Nomura,
  %``Higgs singlet boson as a diphoton resonance in a vectorlike quark model,''
  Phys.\ Rev.\ D {\bf 93}, no. 5, 055034 (2016)
%  doi:10.1103/PhysRevD.93.055034
  [arXiv:1512.06028 [hep-ph]].
  %%CITATION = doi:10.1103/PhysRevD.93.055034;%%
  %119 citations counted in INSPIRE as of 12 Jun 2016
  
  %\cite{Ding:2015rxx}
\bibitem{Ding:2015rxx}
  R.~Ding, L.~Huang, T.~Li and B.~Zhu,
  %``Interpreting $750$ GeV Diphoton Excess with R-parity Violating Supersymmetry,''
  arXiv:1512.06560 [hep-ph].
  %%CITATION = ARXIV:1512.06560;%%
  %123 citations counted in INSPIRE as of 19 Jul 2016
  
  %\cite{Wang:2015kuj}
\bibitem{Wang:2015kuj}
  F.~Wang, L.~Wu, J.~M.~Yang and M.~Zhang,
  %``750 GeV diphoton resonance, 125 GeV Higgs and muon g − 2 anomaly in deflected anomaly mediation SUSY breaking scenarios,''
  Phys.\ Lett.\ B {\bf 759} (2016) 191
%  doi:10.1016/j.physletb.2016.05.071
  [arXiv:1512.06715 [hep-ph]].
  %%CITATION = doi:10.1016/j.physletb.2016.05.071;%%
  %111 citations counted in INSPIRE as of 16 Jun 2016
  
  
  
  %\cite{Dev:2015isx}
\bibitem{Dev:2015isx} 
  P.~S.~B.~Dev and D.~Teresi,
  %``Asymmetric Dark Matter in the Sun and the Diphoton Excess at the LHC,''
  arXiv:1512.07243 [hep-ph].
  %%CITATION = ARXIV:1512.07243;%%
  %116 citations counted in INSPIRE as of 21 Jun 2016

%\cite{Allanach:2015ixl}
\bibitem{Allanach:2015ixl} 
  B.~C.~Allanach, P.~S.~B.~Dev, S.~A.~Renner and K.~Sakurai,
  %``Di-photon Excess Explained by a Resonant Sneutrino in R-parity Violating Supersymmetry,''
  Phys.\ Rev.\ D {\bf 93}, no. 11, 115022 (2016)
%  doi:10.1103/PhysRevD.93.115022
  [arXiv:1512.07645 [hep-ph]].
  %%CITATION = doi:10.1103/PhysRevD.93.115022;%%
  %98 citations counted in INSPIRE as of 21 Jun 2016
  
  %\cite{Cheung:2015cug}
\bibitem{Cheung:2015cug} 
  K.~Cheung, P.~Ko, J.~S.~Lee, J.~Park and P.~Y.~Tseng,
  %``A Higgcision study on the 750 GeV Di-photon Resonance and 125 GeV SM Higgs boson with the Higgs-Singlet Mixing,''
  arXiv:1512.07853 [hep-ph].
  %%CITATION = ARXIV:1512.07853;%%
  %17 citations counted in INSPIRE as of 02 Jan 2016

  
  %\cite{Wang:2015omi}
\bibitem{Wang:2015omi}
  F.~Wang, W.~Wang, L.~Wu, J.~M.~Yang and M.~Zhang,
  %``Interpreting 750 GeV diphoton resonance as degenerate Higgs bosons in NMSSM with vector-like particles,''
  arXiv:1512.08434 [hep-ph].
  %%CITATION = ARXIV:1512.08434;%%
  %85 citations counted in INSPIRE as of 16 Jun 2016
  
  %\cite{Chiang:2015tqz}
\bibitem{Chiang:2015tqz} 
  C.~W.~Chiang, M.~Ibe and T.~T.~Yanagida,
  %``Revisiting Scalar Quark Hidden Sector in Light of 750-GeV Diphoton Resonance,''
  JHEP {\bf 1605}, 084 (2016)
%  doi:10.1007/JHEP05(2016)084
  [arXiv:1512.08895 [hep-ph]].
  %%CITATION = doi:10.1007/JHEP05(2016)084;%%
  %50 citations counted in INSPIRE as of 12 Jun 2016
  
    %\cite{Huang:2015svl}
\bibitem{Huang:2015svl} 
  X.~J.~Huang, W.~H.~Zhang and Y.~F.~Zhou,
  %``A 750 GeV dark matter messenger at the Galactic Center,''
  Phys.\ Rev.\ D {\bf 93}, 115006 (2016)
%  doi:10.1103/PhysRevD.93.115006
  [arXiv:1512.08992 [hep-ph]].
  %%CITATION = doi:10.1103/PhysRevD.93.115006;%%
  %53 citations counted in INSPIRE as of 11 Jul 2016
  
    %\cite{Kanemura:2015bli}
\bibitem{Kanemura:2015bli} 
  S.~Kanemura, K.~Nishiwaki, H.~Okada, Y.~Orikasa, S.~C.~Park and R.~Watanabe,
  %``LHC 750 GeV Diphoton excess in a radiative seesaw model,''
  arXiv:1512.09048 [hep-ph].
  %%CITATION = ARXIV:1512.09048;%%
  
  %\cite{Nomura:2016fzs}
\bibitem{Nomura:2016fzs} 
  T.~Nomura and H.~Okada,
  %``Four-loop Neutrino Model Inspired by Diphoton Excess at 750 GeV,''
  Phys.\ Lett.\ B {\bf 755}, 306 (2016)
  %doi:10.1016/j.physletb.2016.02.022
  [arXiv:1601.00386 [hep-ph]].
  %%CITATION = doi:10.1016/j.physletb.2016.02.022;%%
  %56 citations counted in INSPIRE as of 14 Jul 2016
  
  %\cite{Ko:2016lai}
\bibitem{Ko:2016lai} 
  P.~Ko, Y.~Omura and C.~Yu,
  %``Diphoton Excess at 750 GeV in leptophobic U(1)$^\prime$ model inspired by $E_6$ GUT,''
  JHEP {\bf 1604}, 098 (2016)
%  doi:10.1007/JHEP04(2016)098
  [arXiv:1601.00586 [hep-ph]].
  %%CITATION = doi:10.1007/JHEP04(2016)098;%%
  %56 citations counted in INSPIRE as of 14 Jul 2016
  
  %\cite{Nomura:2016seu}
\bibitem{Nomura:2016seu} 
  T.~Nomura and H.~Okada,
  %``Four-loop Radiative Seesaw Model with 750 GeV Diphoton Resonance,''
  arXiv:1601.04516 [hep-ph].
  %%CITATION = ARXIV:1601.04516;%%
  %44 citations counted in INSPIRE as of 13 Jul 2016
 
 %%%%%%%%%%%%%%%%%%%%%%%%%%%%%%%%%%%%%%%%%%%%%%%%%%%%

%\cite{CMS:2016crm}
\bibitem{CMS:2016crm} 
  CMS Collaboration [CMS Collaboration],
  %``Search for resonant production of high mass photon pairs using $12.9\,\mathrm{fb^{-1}}$ of proton-proton collisions at $\sqrt{s} = 13~\mathrm{TeV}$ and combined interpretation of searches at 8 and 13 TeV,''
  CMS-PAS-EXO-16-027.
  %%CITATION = CMS-PAS-EXO-16-027;%%
  %1 citations counted in INSPIRE as of 12 Aug 2016

%\cite{ATLAS:2016eeo}
\bibitem{ATLAS:2016eeo} 
  The ATLAS collaboration [ATLAS Collaboration],
  %``Search for scalar diphoton resonances with 15.4~fb$^{-1}$ of data collected at $\sqrt{s}$=13 TeV in 2015 and 2016 with the ATLAS detector,''
  ATLAS-CONF-2016-059.
  %%CITATION = ATLAS-CONF-2016-059;%%
  %1 citations counted in INSPIRE as of 12 Aug 2016


%%%%%%%%%%%%%%%%%%%%%%%%%%%%%%%%%%%%%%%%%%%%%%%%%%%%%

  %\cite{Baek:2013qwa}
\bibitem{Baek:2013qwa} 
  S.~Baek, P.~Ko and W.~I.~Park,
  %``Singlet Portal Extensions of the Standard Seesaw Models to a Dark Sector with Local Dark Symmetry,''
  JHEP {\bf 1307}, 013 (2013)
 % doi:10.1007/JHEP07(2013)013
  [arXiv:1303.4280 [hep-ph]].
  %%CITATION = doi:10.1007/JHEP07(2013)013;%%
  
  %\cite{Chiang:2013kqa}
\bibitem{Chiang:2013kqa} 
  C.~W.~Chiang, T.~Nomura and J.~Tandean,
  %``Nonabelian Dark Matter with Resonant Annihilation,''
  JHEP {\bf 1401}, 183 (2014)
%  doi:10.1007/JHEP01(2014)183
  [arXiv:1306.0882 [hep-ph]].
  %%CITATION = doi:10.1007/JHEP01(2014)183;%%
  %12 citations counted in INSPIRE as of 29 Jul 2016

%\cite{Dudas:2013sia}
\bibitem{Dudas:2013sia} 
  E.~Dudas, L.~Heurtier, Y.~Mambrini and B.~Zaldivar,
  %``Extra U(1), effective operators, anomalies and dark matter,''
  JHEP {\bf 1311}, 083 (2013)
%  doi:10.1007/JHEP11(2013)083
  [arXiv:1307.0005 [hep-ph]].
  %%CITATION = doi:10.1007/JHEP11(2013)083;%%
  %17 citations counted in INSPIRE as of 29 Jul 2016
  
  %\cite{Alves:2013tqa}
\bibitem{Alves:2013tqa} 
  A.~Alves, S.~Profumo and F.~S.~Queiroz,
  %``The dark $Z^{'}$ portal: direct, indirect and collider searches,''
  JHEP {\bf 1404}, 063 (2014)
%  doi:10.1007/JHEP04(2014)063
  [arXiv:1312.5281 [hep-ph]].
  %%CITATION = doi:10.1007/JHEP04(2014)063;%%
  %84 citations counted in INSPIRE as of 29 Jul 2016
  
   %\cite{Ko:2014nha}
\bibitem{Ko:2014nha} 
  P.~Ko and Y.~Tang,
  %``Self-interacting scalar dark matter with local $Z_3$ symmetry,''
  JCAP {\bf 1405}, 047 (2014)
%  doi:10.1088/1475-7516/2014/05/047
  [arXiv:1402.6449 [hep-ph], arXiv:1402.6449].
  %%CITATION = doi:10.1088/1475-7516/2014/05/047;%%
 
   %\cite{Ko:2014loa}
\bibitem{Ko:2014loa} 
  P.~Ko and Y.~Tang,
  %``Galactic center $\gamma$-ray excess in hidden sector DM models with dark gauge symmetries: local $Z_{3}$ symmetry as an example,''
  JCAP {\bf 1501}, 023 (2015)
%  doi:10.1088/1475-7516/2015/01/023
  [arXiv:1407.5492 [hep-ph]].
  %%CITATION = doi:10.1088/1475-7516/2015/01/023;%%
  
   %\cite{Baek:2014kna}
\bibitem{Baek:2014kna} 
  S.~Baek, P.~Ko and W.~I.~Park,
  %``Local $Z_2$ scalar dark matter model confronting galactic ${\mathrm GeV}$-scale $\gamma$-ray,''
  Phys.\ Lett.\ B {\bf 747}, 255 (2015)
%  doi:10.1016/j.physletb.2015.06.002
  [arXiv:1407.6588 [hep-ph]].
  %%CITATION = doi:10.1016/j.physletb.2015.06.002;%%
  
  %\cite{Martinez:2014ova}
\bibitem{Martinez:2014ova} 
  R.~Martinez, J.~Nisperuza, F.~Ochoa and J.~P.~Rubio,
  %``Scalar dark matter with CERN-LEP data and $Z′$ search at the LHC in an $U(1)′$ model,''
  Phys.\ Rev.\ D {\bf 90}, no. 9, 095004 (2014)
%  doi:10.1103/PhysRevD.90.095004
  [arXiv:1408.5153 [hep-ph]].
  %%CITATION = doi:10.1103/PhysRevD.90.095004;%%
  %7 citations counted in INSPIRE as of 29 Jul 2016
  
  %\cite{Martinez:2014rea}
\bibitem{Martinez:2014rea} 
  R.~Martinez, J.~Nisperuza, F.~Ochoa, J.~P.~Rubio and C.~F.~Sierra,
  %``Scalar coupling limits and diphoton Higgs decay from LHC in an U(1)′ model with scalar dark matter,''
  Phys.\ Rev.\ D {\bf 92}, no. 3, 035016 (2015)
%  doi:10.1103/PhysRevD.92.035016
  [arXiv:1411.1641 [hep-ph]].
  %%CITATION = doi:10.1103/PhysRevD.92.035016;%%
  %2 citations counted in INSPIRE as of 29 Jul 2016
  
  %\cite{Alves:2015pea}
\bibitem{Alves:2015pea} 
  A.~Alves, A.~Berlin, S.~Profumo and F.~S.~Queiroz,
  %``Dark Matter Complementarity and the Z$^\prime$ Portal,''
  Phys.\ Rev.\ D {\bf 92}, no. 8, 083004 (2015)
%  doi:10.1103/PhysRevD.92.083004
  [arXiv:1501.03490 [hep-ph]].
  %%CITATION = doi:10.1103/PhysRevD.92.083004;%%
  %42 citations counted in INSPIRE as of 29 Jul 2016
  
    %\cite{Guo:2015lxa}
\bibitem{Guo:2015lxa} 
  J.~Guo, Z.~Kang, P.~Ko and Y.~Orikasa,
  %``Accidental dark matter: Case in the scale invariant local B-L model,''
  Phys.\ Rev.\ D {\bf 91}, no. 11, 115017 (2015)
%  doi:10.1103/PhysRevD.91.115017
  [arXiv:1502.00508 [hep-ph]].
  %%CITATION = doi:10.1103/PhysRevD.91.115017;%%
  
  %\cite{Alves:2015mua}
\bibitem{Alves:2015mua} 
  A.~Alves, A.~Berlin, S.~Profumo and F.~S.~Queiroz,
  %``Dirac-fermionic dark matter in U(1)$_{X}$ models,''
  JHEP {\bf 1510}, 076 (2015)
%  doi:10.1007/JHEP10(2015)076
  [arXiv:1506.06767 [hep-ph]].
  %%CITATION = doi:10.1007/JHEP10(2015)076;%%
  %54 citations counted in INSPIRE as of 29 Jul 2016

%\cite{Ducu:2015fda}
\bibitem{Ducu:2015fda} 
  O.~Ducu, L.~Heurtier and J.~Maurer,
  %``LHC signatures of a Z' mediator between dark matter and the SU(3) sector,''
  JHEP {\bf 1603}, 006 (2016)
%  doi:10.1007/JHEP03(2016)006
  [arXiv:1509.05615 [hep-ph]].
  %%CITATION = doi:10.1007/JHEP03(2016)006;%%

%\cite{Martinez:2015wrp}
\bibitem{Martinez:2015wrp} 
  R.~Martinez and F.~Ochoa,
  %``Spin-independent interferences and spin-dependent interactions with scalar dark matter,''
  JHEP {\bf 1605}, 113 (2016)
%  doi:10.1007/JHEP05(2016)113
  [arXiv:1512.04128 [hep-ph]].
  %%CITATION = doi:10.1007/JHEP05(2016)113;%%


%%%%%%%%%%%%%%%%%%%%%%%%%%%%%%%%%%%%%%%%%%%%%%%%%%%%%%

 %\cite{Chpoi:2013wga}
\bibitem{Chpoi:2013wga} 
  S.~Choi, S.~Jung and P.~Ko,
  %``Implications of LHC data on 125 GeV Higgs-like boson for the Standard Model and its various extensions,''
  JHEP {\bf 1310}, 225 (2013)
%doi:10.1007/JHEP10(2013)225
  [arXiv:1307.3948 [hep-ph]];
  %%CITATION = doi:10.1007/JHEP10(2013)225;%%
%\cite{Cheung:2015dta}
\bibitem{Cheung:2015dta} 
  K.~Cheung, P.~Ko, J.~S.~Lee and P.~Y.~Tseng,
  %``Bounds on Higgs-Portal models from the LHC Higgs data,''
  JHEP {\bf 1510}, 057 (2015)
%  doi:10.1007/JHEP10(2015)057
  [arXiv:1507.06158 [hep-ph]];
  %%CITATION = doi:10.1007/JHEP10(2015)057;%%
%\cite{Cheung:2015cug}

%\cite{Hook:2010tw}
\bibitem{Hook:2010tw} 
  A.~Hook, E.~Izaguirre and J.~G.~Wacker,
  %``Model Independent Bounds on Kinetic Mixing,''
  Adv.\ High Energy Phys.\  {\bf 2011}, 859762 (2011)
%  doi:10.1155/2011/859762
  [arXiv:1006.0973 [hep-ph]].
  %%CITATION = doi:10.1155/2011/859762;%%
  %91 citations counted in INSPIRE as of 05 Nov 2016

 %\cite{Andreas:2012mt}
\bibitem{Andreas:2012mt}
  S.~Andreas, C.~Niebuhr and A.~Ringwald,
  %``New Limits on Hidden Photons from Past Electron Beam Dumps,''
  Phys.\ Rev.\ D {\bf 86} (2012) 095019
%  doi:10.1103/PhysRevD.86.095019
  [arXiv:1209.6083 [hep-ph]].
  %%CITATION = doi:10.1103/PhysRevD.86.095019;%%
  %83 citations counted in INSPIRE as of 23 Jun 2016
  
  %\cite{Jaeckel:2012yz}
\bibitem{Jaeckel:2012yz} 
  J.~Jaeckel, M.~Jankowiak and M.~Spannowsky,
  %``LHC probes the hidden sector,''
  Phys.\ Dark Univ.\  {\bf 2}, 111 (2013)
%  doi:10.1016/j.dark.2013.06.001
  [arXiv:1212.3620 [hep-ph]].
  %%CITATION = doi:10.1016/j.dark.2013.06.001;%%
  %64 citations counted in INSPIRE as of 23 Jun 2016

   \bibitem{Gunion:1989we} 
  J.~F.~Gunion, H.~E.~Haber, G.~L.~Kane and S.~Dawson,
  %``The Higgs Hunter's Guide,''
  Front.\ Phys.\  {\bf 80}, 1 (2000).
  %%CITATION = FRPHA,80,1;%%
  %396 citations counted in INSPIRE as of 12 Jul 2015
  
  %\cite{Ibarra:2014qma}
\bibitem{Ibarra:2014qma} 
  A.~Ibarra, T.~Toma, M.~Totzauer and S.~Wild,
  %``Sharp Gamma-ray Spectral Features from Scalar Dark Matter Annihilations,''
  Phys.\ Rev.\ D {\bf 90}, no. 4, 043526 (2014)
%  doi:10.1103/PhysRevD.90.043526
  [arXiv:1405.6917 [hep-ph]].
  %%CITATION = doi:10.1103/PhysRevD.90.043526;%%
  %6 citations counted in INSPIRE as of 15 Jun 2016
  
  %\cite{Giacchino:2015hvk}
\bibitem{Giacchino:2015hvk} 
  F.~Giacchino, A.~Ibarra, L.~L.~Honorez, M.~H.~G.~Tytgat and S.~Wild,
  %``Signatures from Scalar Dark Matter with a Vector-like Quark Mediator,''
  JCAP {\bf 1602}, no. 02, 002 (2016)
%  doi:10.1088/1475-7516/2016/02/002
  [arXiv:1511.04452 [hep-ph]].
  %%CITATION = doi:10.1088/1475-7516/2016/02/002;%%
  %7 citations counted in INSPIRE as of 15 Jun 2016
  
  %\cite{Belanger:2014vza}
\bibitem{Belanger:2014vza} 
  G.~Belanger, F.~Boudjema, A.~Pukhov and A.~Semenov,
  %``micrOMEGAs4.1: two dark matter candidates,''
  Comput.\ Phys.\ Commun.\  {\bf 192}, 322 (2015)
%  doi:10.1016/j.cpc.2015.03.003
  [arXiv:1407.6129 [hep-ph]].
  %%CITATION = doi:10.1016/j.cpc.2015.03.003;%%
  %84 citations counted in INSPIRE as of 07 Jul 2016
  
    %\cite{Gondolo:1990dk}
\bibitem{Gondolo:1990dk} 
  P.~Gondolo and G.~Gelmini,
  %``Cosmic abundances of stable particles: Improved analysis,''
  Nucl.\ Phys.\ B {\bf 360}, 145 (1991).
 % doi:10.1016/0550-3213(91)90438-4
  %%CITATION = doi:10.1016/0550-3213(91)90438-4;%%
  %648 citations counted in INSPIRE as of 15 Jan 2016
  
    %\cite{Ade:2013zuv}
\bibitem{Ade:2013zuv} 
  P.~A.~R.~Ade {\it et al.}  [Planck Collaboration],
  %``Planck 2013 results. XVI. Cosmological parameters,''
  Astron.\ Astrophys.\  (2014)
  [arXiv:1303.5076 [astro-ph.CO]].
  %%CITATION = ARXIV:1303.5076;%%
  %2596 citations counted in INSPIRE as of 08 Nov 2014
  
  
  %\cite{Hisano:2015bma}
\bibitem{Hisano:2015bma} 
  J.~Hisano, R.~Nagai and N.~Nagata,
  %``Effective Theories for Dark Matter Nucleon Scattering,''
  JHEP {\bf 1505}, 037 (2015)
%  doi:10.1007/JHEP05(2015)037
  [arXiv:1502.02244 [hep-ph]].
  %%CITATION = doi:10.1007/JHEP05(2015)037;%%
  %9 citations counted in INSPIRE as of 04 Apr 2016
  
  %\cite{Akerib:2013tjd}
\bibitem{Akerib:2013tjd} 
  D.~S.~Akerib {\it et al.} [LUX Collaboration],
  %``First results from the LUX dark matter experiment at the Sanford Underground Research Facility,''
  Phys.\ Rev.\ Lett.\  {\bf 112}, 091303 (2014)
%  doi:10.1103/PhysRevLett.112.091303
  [arXiv:1310.8214 [astro-ph.CO]].
  %%CITATION = doi:10.1103/PhysRevLett.112.091303;%%
  %1309 citations counted in INSPIRE as of 13 Jul 2016
  
  %\cite{Akerib:2015rjg}
\bibitem{Akerib:2015rjg} 
  D.~S.~Akerib {\it et al.} [LUX Collaboration],
  %``Improved Limits on Scattering of Weakly Interacting Massive Particles from Reanalysis of 2013 LUX Data,''
  Phys.\ Rev.\ Lett.\  {\bf 116}, no. 16, 161301 (2016)
%  doi:10.1103/PhysRevLett.116.161301
  [arXiv:1512.03506 [astro-ph.CO]].
  %%CITATION = doi:10.1103/PhysRevLett.116.161301;%%
  %112 citations counted in INSPIRE as of 13 Jul 2016
  
  %\cite{Aprile:2015uzo}
\bibitem{Aprile:2015uzo} 
  E.~Aprile {\it et al.} [XENON Collaboration],
  %``Physics reach of the XENON1T dark matter experiment,''
  JCAP {\bf 1604}, no. 04, 027 (2016)
%  doi:10.1088/1475-7516/2016/04/027
  [arXiv:1512.07501 [physics.ins-det]].
  %%CITATION = doi:10.1088/1475-7516/2016/04/027;%%
  %41 citations counted in INSPIRE as of 13 Jul 2016

%\cite{Ackermann:2015zua}
\bibitem{Ackermann:2015zua} 
  M.~Ackermann {\it et al.} [Fermi-LAT Collaboration],
  %``Searching for Dark Matter Annihilation from Milky Way Dwarf Spheroidal Galaxies with Six Years of Fermi Large Area Telescope Data,''
  Phys.\ Rev.\ Lett.\  {\bf 115}, no. 23, 231301 (2015)
%  doi:10.1103/PhysRevLett.115.231301
  [arXiv:1503.02641 [astro-ph.HE]].
  %%CITATION = doi:10.1103/PhysRevLett.115.231301;%%
  %249 citations counted in INSPIRE as of 07 Jul 2016
 
%\cite{Khachatryan:2014rra}
\bibitem{Khachatryan:2014rra} 
  V.~Khachatryan {\it et al.} [CMS Collaboration],
  %``Search for dark matter, extra dimensions, and unparticles in monojet events in proton\UTF{2013}proton collisions at $\sqrt{s} = 8$ TeV,''
  Eur.\ Phys.\ J.\ C {\bf 75}, no. 5, 235 (2015)
%  doi:10.1140/epjc/s10052-015-3451-4
  [arXiv:1408.3583 [hep-ex]].
  %%CITATION = doi:10.1140/epjc/s10052-015-3451-4;%%
  %225 citations counted in INSPIRE as of 11 Jul 2016

%\cite{Aad:2014aqa}
\bibitem{Aad:2014aqa} 
  G.~Aad {\it et al.} [ATLAS Collaboration],
  %``Search for new phenomena in the dijet mass distribution using $p-p$ collision data at $\sqrt{s}=8$ TeV with the ATLAS detector,''
  Phys.\ Rev.\ D {\bf 91}, no. 5, 052007 (2015)
 % doi:10.1103/PhysRevD.91.052007
  [arXiv:1407.1376 [hep-ex]].
  %%CITATION = doi:10.1103/PhysRevD.91.052007;%%
  %255 citations counted in INSPIRE as of 04 Nov 2016
  
   %\cite{Khachatryan:2015dcf}
\bibitem{Khachatryan:2015dcf} 
  V.~Khachatryan {\it et al.} [CMS Collaboration],
  %``Search for narrow resonances decaying to dijets in proton-proton collisions at $\sqrt(s) =$ 13 TeV,''
  Phys.\ Rev.\ Lett.\  {\bf 116}, no. 7, 071801 (2016)
%  doi:10.1103/PhysRevLett.116.071801
  [arXiv:1512.01224 [hep-ex]].
  %%CITATION = doi:10.1103/PhysRevLett.116.071801;%%
  %51 citations counted in INSPIRE as of 28 Apr 2016
 
 %\cite{ATLAS:2015nsi}
\bibitem{ATLAS:2015nsi} 
  G.~Aad {\it et al.} [ATLAS Collaboration],
  %``Search for new phenomena in dijet mass and angular distributions from $pp$ collisions at $\sqrt{s}=$ 13 TeV with the ATLAS detector,''
  Phys.\ Lett.\ B {\bf 754}, 302 (2016)
%  doi:10.1016/j.physletb.2016.01.032
  [arXiv:1512.01530 [hep-ex]].
  %%CITATION = doi:10.1016/j.physletb.2016.01.032;%%
  %44 citations counted in INSPIRE as of 28 Apr 2016
  
  

\end{thebibliography}
\end{document}